\documentclass[onecolumn]{mn2e}
\usepackage{graphicx}
\begin{document}
\title[Rapidly rotating superfluid neutron stars in Newtonian dynamics]{ 
Rapidly rotating superfluid neutron stars in Newtonian dynamics}
\author[S. Yoshida and Y. Eriguchi]{Shijun Yoshida$^1$\thanks{E-mail: yoshida@fisica.ist.utl.pt} and 
Yoshiharu Eriguchi$^2$\thanks{E-mail: eriguchi@esa.c.u-tokyo.ac.jp} \\
$^1$Centro Multidisciplinar de Astrof\'{\i}sica -- CENTRA,
           Departamento de F\'{\i}sica, Instituto Superior T\'ecnico, \\
           Av. Rovisco Pais 1, 1049-001 Lisboa, Portugal \\
$^2$Department of Earth Science and Astronomy, Graduate School of Arts and Sciences, 
           University of Tokyo, \\ 
           Komaba, Meguro, Tokyo 153-8902, Japan 
}
\date{Typeset \today ; Received / Accepted}
\maketitle
\begin{abstract}
We develop a formulation for constructing and examining rapidly rotating 
Newtonian neutron star models that contain two superfluids, taking account 
of the effect of the rotation velocity difference between two superfluids. 
We assume neutron stars to be composed of the superfluid neutrons and the 
mixture of the superfluid protons and the normal fluid electrons. To describe 
Newtonian dynamics of the two superfluids, the Newtonian version of the 
so-called two-fluid formalism is employed. The effect of the rotation 
velocity difference on the structure of equilibrium state is treated as a 
small perturbation to rapidly rotating superfluid stars whose angular 
velocities of two superfluids are assumed to be exactly the same. We derive 
basic equations for the perturbed structures of rapidly rotating superfluid 
stars due to the rotation velocity difference between two superfluids. 
Assuming the superfluids to obey a simple analytical equation of state 
proposed by Prix, Comer, and Andersson, we obtain numerical solutions for the 
perturbations and find that the density distributions of the superfluids are 
strongly dependent on the parameter $\sigma$ which appears in the analytical 
equation of state and characterizes the so-called symmetry energy.
It is also found that if Prix et al.'s analytical equation of state is assumed, 
the perturbations can be represented in terms of the universal functions that 
are independent of the parameters of the equation of state. 
\end{abstract} 

\section{Introduction}

To investigate properties of equilibrium configurations of rotating neutron 
stars, so far, most neutron star models have been obtained by assuming neutron 
star matter to be a one-constituent perfect fluid (for a review, see, e.g. 
Stergioulas 2003). This treatment of equilibrium states of neutron stars seems 
to be quite reasonable as a first approximation for examining global 
properties of neutron stars such as the gravitational mass, the radius, or the 
maximum rotation frequency.  However, it has long been suggested that neutrons 
in the inner crust and neutrons and protons in the core of neutron stars are 
in superfluid states when the interior temperatures cool down below 
$T_c\sim 10^9{\rm K}$ (e.g., Shapiro \& Teukolsky 1983). Since the 
interior temperature of neutron stars is believed to cool down quickly via the 
neutrino emission (e.g., Baym \& Pethick 1979), it is likely that each of many 
observable neutron stars, except newly born ones, has a core containing 
superfluids. Although superfluidity in the interior might be one of the 
important ingredients that affect the structures of neutron stars, the 
superfluidity has been neglected in most studies on equilibrium 
configurations of rotating neutron stars. Thus, it is necessary to 
examine how superfluidity affects fundamental properties of rotating neutron 
stars.

The superfluidity in neutron stars has been mainly argued in connection with 
the glitch phenomenon and the post-glitch relaxation observed in pulsars. The 
glitch is a sudden decrease of an observed pulse period $T_p$ of a pulsar, 
whose largest magnitude is $\Delta T_p/T_p\sim -10^{-6}$, that is followed 
by a period of continuous relaxation. In most explanations of glitch 
phenomena, a two-component model has been employed to describe rotational 
dynamics of neutron stars containing superfluid neutrons. In those treatments, 
one of the components is a mixture of the charged particles such as protons 
(or ions) and electrons, which are assumed to co-rotate because the charged 
components are strongly coupled each other due to short-range electromagnetic 
interactions (Easson 1979, Alpar et al. 1984), and the other the superfluid 
neutrons, which are supposed to rotate separately from the charged components 
because the superfluid neutrons are inviscous and loosely couple to the 
charged components. In this paper, we call the mixture of the charged 
components ``protons'' for brevity. It has been shown that those two-component 
superfluid models have succeeded in explaining observed features of the 
glitches and the post-glitch relaxations qualitatively (Anderson \& Itoh 1975; 
Alpar et al. 1984; Sedrakian et al. 1995, Link \& Epstein 1996).  Those 
results mean that the existence of superfluid components inside neutron stars 
is supported not only from the theory of nuclear physics but also from 
observations of pulsars.

If they co-rotate, as mentioned in Prix \& Rieutord (2002), two constituents 
in a rotating superfluid star behave as one-constituent ordinary fluid, 
because the so-called entrainment effect, due to which the momentum of one 
superfluid constituent is dependent on the mass current of all superfluid 
constituents, does not operate and two constituents are in a chemical 
equilibrium state. On the other hand, specific characteristics of neutron 
stars due to the superfluidity are considered to be observed when two 
constituents rotate at different rotation rates. To study the effect of 
rotation velocity difference between two constituents on the global structures 
of rotating neutron stars with superfluidity, Prix (1999) derived basic 
equations for the equilibrium configurations and obtained analytical solutions 
for slowly rotating superfluid stars within the framework of Newtonian 
dynamics, generalizing the so-called Chandrasekhar-Milne expansion for 
ordinary fluid stars (see, e.g. Tassoul 1978), in which rotational effects 
are treated as perturbations to non-rotating spherical stars. To describe 
dynamics of superfluids, he made use of a variant of the so-called two-fluid 
formalism, which had been mainly developed by Carter, Langlois, and their 
co-workers (Carter 1989; Carter \& Langlois 1998; Langlois, Sedrakian, \& 
Carter 1998; Prix 2002). In order to obtain analytical solutions, Prix (1999) 
employed an equation of state whose functional form is $P\propto\rho^2$, 
where $P$ and $\rho$ are the pressure and density, respectively.  More 
recently, Prix, Comer, \& Andersson (2002a) extended the Prix formalism so as 
to include the entrainment effect between two superfluids in order to 
investigate its effect on the stellar structures.  In the same paper, 
Prix et al. (2002a) also argued the effect of the so-called symmetry energy, 
which can be included as a parameter in the equation of state, and found that 
the effect of symmetry energy is significant in determination of the density 
distributions of the neutrons and the protons.  As for relativistic rotating 
superfluid stars, Andersson \& Comer (2001) calculated equilibrium 
configurations of slowly rotating superfluid neutron stars, extending the 
slow rotation approximation formalism devised by Hartle (1967) (see, also 
Hartle \& Thorne 1968) for ordinary fluid stars to relativistic 
two-constituent superfluid stars. Very recently, Prix, Novak, \& Comer (2002b) 
have obtained preliminary results for rapidly rotating superfluid stars, 
without the slow rotation approximation.

The purpose of this paper is to improve our understanding of properties of 
equilibrium configurations of rapidly rotating neutron stars with superfluids. 
We calculate equilibrium configurations of rapidly rotating neutron stars with 
neutron and proton superfluids.  In this paper, we are concerned with rotating
neutron stars of two superfluids whose angular velocities are different.  We 
however assume that the rotation velocity difference between two superfluids 
is very small in comparison with the angular velocity of the star. This 
assumption could be reasonable for models of superfluid neutron stars because 
observations of pulsar glitches show that the rotation velocity of the protons 
may differ from that of the neutrons, but the amount of the rotation velocity 
difference is not very large. 
We can therefore treat the effect of the rotation velocity difference between 
two superfluids on the structures of rapidly rotating neutron stars as a small 
perturbation to equilibrium configurations of rapidly rotating neutron stars 
in which two superfluids co-rotate. For the dynamics of superfluids inside 
stars, we employ a Newtonian version of the two-constituent formalism 
developed by Prix (2002). In order to obtain basic equations for determining 
the effect of the rotation velocity difference between two superfluids, 
equations of superfluid hydrostatic equilibrium and the Poisson equation are 
expanded in terms of the rotation velocity difference between two superfluids. 
To obtain the numerical solutions of equilibrium configurations, we make use 
of a variant of the so-called  Self Consistent Field method (SCF) devised by 
Ostriker \& Mark (1968). The equation of state for the superfluid we use is 
the analytical one employed by Prix et al. (2002a), which is a natural 
extension of an $N=1$ polytropic equation of state for a barotropic 
ordinary fluid to the case of two superfluids.  In \S 2 we present the basic 
equations employed in this paper for the rapidly rotating superfluid stars. 
Numerical results are given in \S 3, and \S 4 is devoted for summary and
discussion.

\section{Formulation}

\subsection{Two-constituent formalism for Newtonian superfluid dynamics}

In this paper, a neutron star is assumed to be composed of superfluid neutrons 
and a mixture of superfluid protons and normal fluid electrons, because the 
interior temperatures of old neutron stars are much lower than the transition 
temperature to neutron and proton superfluids, which is considered to be 
$T_c\sim 10^9$ K (see, e.g., Epstein 1988). We further assume that this 
mixture of charged particles is perfectly in the state of charge neutrality 
because of strong coupling between the protons and the electrons by 
electromagnetic interactions. The electrons therefore co-move with the 
protons, and their motions can be described with one fluid velocity 
$v^a_p$. Through this paper, we call the mixture of the the protons and the 
electrons ``protons'' for brevity as mentioned before. To describe the 
dynamics of the superfluids, we make use of the two-constituent formalism 
for Newtonian superfluid dynamics, which has been recently developed by 
Prix (2002).

We take account of no ``transfusion'' between the neutrons and the protons. 
The particle numbers of the neutron and the proton must be therefore conserved 
separately. The particle number conservation equations for the neutron and 
the proton are then given by 
\begin{equation}
\nabla_a(n_nv_n^a) = 0\,,\quad \nabla_a(n_pv_p^a) = 0\,, 
\end{equation} 
where $n_n$ and $n_p$ are the number densities of the neutron and the proton, 
and $v_n^a$ and $v_p^a$ denote the fluid velocities of the neutron and the 
proton, respectively. Here, $\nabla_a$ means the covariant derivative in the 
three dimensional flat space. The mass densities of two superfluids can be 
written as
\begin{equation}
\rho_n = m n_n\,,\quad \rho_p = m n_p\,,
\end{equation}
where $m$ is the mass of the neutron, and we assume the mass of the proton to 
be equal to that of the neutron. The fundamental quantity of the 
two-constituent formalism for the Newtonian superfluid dynamics is the 
internal energy density $\cal E$, which is a function of $\rho_n$, $\rho_p$, 
and $\Delta^2$, where $\Delta^2=\Delta^a \Delta_a$ and $\Delta^a=v_p^a-v_n^a$. 
Note that we have neglected the entropy $s$ carried by the normal fluid of the 
electron for simplicity. This treatment is justified for neutron stars whose 
internal temperatures are sufficiently low. This function $\cal E$ defines 
several basic thermodynamical quantities which describe the dynamics of two 
superfluids in terms of the total difference as follows
\begin{equation}
d{\cal E}=\tilde\mu_n\,d\rho_n+\tilde\mu_p\,d\rho_p+\alpha\,d\Delta^2\,,
\end{equation}
where $\tilde\mu_n$ and $\tilde\mu_p$ denote the specific chemical potentials 
of the neutron and the proton, respectively. The function $\alpha$ represents 
the strength of the entrainment effect. By using a variant of the Gibbs-Duhem 
relation, the generalized pressure of two fluids is defined by 
\begin{equation}
dP=\rho_n\,d\tilde\mu_n+\rho_p\,d\tilde\mu_p-\alpha\,d\Delta^2\,,
\end{equation}
and we can obtain the relation,  
\begin{equation}
P=\rho_n\,\tilde\mu_n+\rho_p\,\tilde\mu_p-{\cal E}\,.
\end{equation}
According to Prix (2002), the Euler equations for two superfluids can be given 
by 
\begin{equation}
(\partial_t+{\cal L}_{v_n})(v_{n,a}+\varepsilon_n\Delta_a)+
\nabla_a\left(\tilde\mu_n+\Phi-{1\over 2}\,|v_n|^2\right)=0\,,
\label{euler1}
\end{equation}
\begin{equation}
(\partial_t+{\cal L}_{v_p})(v_{p,a}-\varepsilon_p\Delta_a)+
\nabla_a\left(\tilde\mu_p+\Phi-{1\over 2}\,|v_p|^2\right)=0\,,
\label{euler2}
\end{equation}
where $\varepsilon_n=2\alpha/\rho_n$ and $\varepsilon_p=2\alpha/\rho_p$. Here, 
${\cal L}_u$ means the Lie derivative along the vector field $u^a$. Note that 
we have considered no direct interaction force between two superfluids such as
the mutual friction or the Magnus-type force. In equations (\ref{euler1}) and 
(\ref{euler2}), the function $\Phi$ is the gravitational potential, which is 
determined with the total mass density $\rho$, given by $\rho=\rho_n+\rho_p$, 
through the Poisson equation 
\begin{equation}
\nabla_a\nabla^a\Phi=4\pi G\rho\,,
\label{poisson}
\end{equation}
where $G$ is the gravitational constant. Considering the exterior 
differentiation of equations (\ref{euler1}) and (\ref{euler2}), we can obtain 
the vorticity equations, given by
\begin{equation}
(\partial_t+{\cal L}_{v_n})\omega_{n,ab}=0\,, \quad 
(\partial_t+{\cal L}_{v_p})\omega_{p,ab}=0\,,
\label{vorticity}
\end{equation}
where $\omega_{n,ab}$ and $\omega_{p,ab}$ are, 
respectively, the exterior derivatives of the one-forms $v_{n,a}+\varepsilon_n\Delta_a$ and 
$v_{p,a}-\varepsilon_p\Delta_a$, which are defined by 
\begin{equation}
\omega_{n,ab}=\partial_a(v_{n,b}+\varepsilon_n\Delta_b)-
\partial_b(v_{n,a}+\varepsilon_n\Delta_a)\,,\quad 
\omega_{p,ab}=\partial_a(v_{p,b}-\varepsilon_p\Delta_b)-
\partial_b(v_{p,a}-\varepsilon_p\Delta_a)\,.
\end{equation}

\subsection{Stationary and axisymmetric equilibrium configurations}

We consider stationary and axisymmetric equilibrium configurations. Thus, the 
time and azimuthal derivatives of physical quantities must vanish because a 
star is assumed to be in a stationary and axisymmetric state. In this paper,
we further assume that the neutrons and the protons, respectively, rotate with 
the angular velocities $\Omega_n$ and $\Omega_p$ around the axis of rotation, 
and that no meridional circulation is present in a superfluid star. The fluid 
velocities for two fluids are then given by 
\begin{equation}
v_n^a=\Omega_n \varphi^a\,,\quad v_p^a=\Omega_p \varphi^a\,,
\label{fluid_velocity}
\end{equation}
where $\varphi^a$ is the rotational Killing vector. In this paper, we employ 
the spherical polar coordinates $(r,\theta,\varphi)$. The components 
of $\varphi^a$ can be written as $\varphi^a=\delta^a_\varphi$ in the 
spherical polar coordinates. First of all, let us consider the vorticity 
conservation for the flow given by equation (\ref{fluid_velocity}). 
Substituting equation (\ref{fluid_velocity}) into equation (\ref{vorticity}), 
we obtain 
\begin{equation}
({\cal L}_{v_n}\omega_n)_{ab}\,dx^a\wedge dx^b=d[\varpi^2 
\lbrace(1-\varepsilon_n)\Omega_n+\varepsilon_n\Omega_p \rbrace]\wedge d\Omega_n=0\,,
\end{equation}
\begin{equation}
({\cal L}_{v_p}\omega_p)_{ab}\,dx^a\wedge dx^b=d[\varpi^2
\lbrace(1-\varepsilon_p)\Omega_p+\varepsilon_p\Omega_n \rbrace]\wedge d\Omega_p=0\,,
\end{equation}
where $\varpi$ is the distance from the rotation axis, defined by 
$\varpi=r\cos\theta$. Thus, these vorticity conservation equations are 
automatically satisfied if the rotation laws for two superfluids are given 
as follows: 
\begin{equation}
\varpi^2 \lbrace(1-\varepsilon_n)\Omega_n+\varepsilon_n\Omega_p \rbrace = j_n(\Omega_n)\,,\quad 
\varpi^2 \lbrace(1-\varepsilon_p)\Omega_p+\varepsilon_p\Omega_n \rbrace = j_p(\Omega_p)\,, 
\end{equation}
where $j_n(\Omega_n)$ and $j_p(\Omega_p)$ are arbitrary functions of 
$\Omega_n$ and $\Omega_p$, respectively. In other words, the rotation laws of 
superfluid stars must be strongly restricted because the rotation velocities 
$\Omega_n$ and $\Omega_p$ may depend on the entrainment functions 
$\varepsilon_n$ and $\varepsilon_p$, whose functional forms are in general 
determined with the matter distributions of stars.  When the effect of the 
entrainment between two superfluids does not operate, that is, the case of 
$\alpha=0$ or of $\Omega_n=\Omega_p$, the vorticity conservation equations 
become relatively simple as follows
\begin{equation}
d\varpi\wedge d\Omega_n=0\,,\quad 
d\varpi\wedge d\Omega_p=0\,. 
\end{equation}
The functions $\Omega_n$ and $\Omega_p$ must not therefore depend on 
$z=r\sin\theta$ in this situation. In other words, we can choose the rotation 
velocities $\Omega_n$ and $\Omega_p$ freely as long as 
$\Omega_n=\Omega_n(\varpi)$ and $\Omega_p=\Omega_p(\varpi)$ are fulfilled. 
Note that this condition is the same as that of barotropic ordinary fluid 
stars.  For simplicity, in this paper, we assume that both the neutrons and 
the protons are uniformly rotating, i.e. $\Omega_n={\rm const}$ and 
$\Omega_p={\rm const}$, because vorticity conservations are automatically 
satisfied for uniformly rotating configurations. 
Assuming the uniform rotations and integrating the Euler equations 
(\ref{euler1}) and (\ref{euler2}), we can obtain integrated equations for 
hydrostatic equilibrium states, given by
\begin{equation}
\tilde\mu_n+\Phi-{\varpi^2\over 2}\,\Omega_n^2=C_n\,,\quad 
\tilde\mu_p+\Phi-{\varpi^2\over 2}\,\Omega_p^2=C_p\,,
\label{beq1}
\end{equation} 
where $C_n$ and $C_p$ are integral constants. Here, we have used the 
relationships 
\begin{eqnarray}
{\cal L}_{v_n}(v_n+\varepsilon_n\Delta)_adx^a&=&\varpi^2
\lbrace(1-\varepsilon_n)\Omega_n+\varepsilon_n\Omega_p\rbrace\,d\Omega_n=0\,,\nonumber \\
{\cal L}_{v_p}(v_p-\varepsilon_p\Delta)_adx^a&=&\varpi^2
\lbrace(1-\varepsilon_p)\Omega_p+\varepsilon_p\Omega_n\rbrace\,d\Omega_p=0\,.
\end{eqnarray}

We are interested in configurations in which two superfluids are rotating 
rapidly with different angular velocities, $\Omega_n\ne\Omega_p$. We do 
not therefore make use of any slow-rotation approximation such as the 
Chandrasekhar-Milne expansion. We however assume the difference between 
rotation velocities of two superfluids to be very small. We can then treat the 
effect of the angular velocity difference $\Omega_p-\Omega_n$ on the stellar 
structure as a perturbation to rapidly rotating stars whose rotation rates 
of two superfluids are exactly the same. For superfluid neutron stars, 
although the angular velocity $\Omega_n$ may differ from the angular 
velocity $\Omega_p$ because the interaction between two superfluids may 
be very weak due to the superfluidity, it is believed that this difference 
between two angular velocities is small in comparison with the averaged 
angular velocity of the star. Thus, this treatment could be appropriate
for neutron star models with superfluidity.

We expand physical quantities appeared in equations (\ref{poisson}) and 
(\ref{beq1}) up to the first order in 
$(\Omega_n-\Omega_p)/(|\Omega_n|+|\Omega_p|)$ as follows:
\begin{eqnarray}
\Omega_n=\Omega(1+\delta\Omega_n)\,,\quad 
\Omega_p=\Omega(1+\delta\Omega_p)\,,
\end{eqnarray} 
\begin{eqnarray}
\rho_n=\rho_{n0}+\delta\rho_n\,,\quad 
\rho_p=\rho_{p0}+\delta\rho_p\,,
\end{eqnarray}
\begin{eqnarray}
\tilde\mu_n=\tilde\mu_{n0}+\delta\tilde\mu_n\,,\quad 
\tilde\mu_p=\tilde\mu_{p0}+\delta\tilde\mu_p\,,
\end{eqnarray}
\begin{eqnarray}
\Phi=\Phi_0+\delta\Phi\,,\quad C_n=C_{n0}+\delta C_n\,,\quad 
C_p=C_{p0}+\delta C_p\,,
\end{eqnarray}
where $\tilde\mu_{n0}$, $\tilde\mu_{p0}$, $\delta\tilde\mu_n$, and 
$\delta\tilde\mu_p$ are given in terms of partial derivatives of the internal 
energy density $\cal E$, $\delta\rho_n$, and $\delta\rho_p$ by
\begin{eqnarray}
\tilde\mu_{n0}={\partial{\cal E}\over\partial\rho_n}\,,\quad
\tilde\mu_{p0}={\partial{\cal E}\over\partial\rho_p}\,,
\end{eqnarray}
\begin{eqnarray}
\delta\tilde\mu_n={\partial^2{\cal E}\over\partial\rho_n^2}\delta\rho_n+
{\partial^2{\cal E}\over\partial\rho_n\partial\rho_p}\delta\rho_p\,,\quad
\delta\tilde\mu_p={\partial^2{\cal E}\over\partial\rho_p^2}\delta\rho_p+
{\partial^2{\cal E}\over\partial\rho_p\partial\rho_n}\delta\rho_n\,. 
\end{eqnarray}
Here, quantities following ``$\delta$'' denote perturbed ones. 
Note that since we consider the accuracy up to the first order in the rotation 
velocity difference, no entrainment effects appear in equations we solve. 
Substituting those expanded quantities into equations (\ref{poisson}) and 
(\ref{beq1}), we obtain the zero-th order equations, given by
\begin{eqnarray}
\tilde\mu_{n0}+\Phi_0-{\varpi^2\over 2}\Omega^2=C_{n0}\,,
\label{eq0-1}
\end{eqnarray}
\begin{eqnarray}
\tilde\mu_{p0}+\Phi_0-{\varpi^2\over 2}\Omega^2=C_{p0}\,,
\label{eq0-2}
\end{eqnarray}
\begin{eqnarray}
\nabla_a\nabla^a\Phi_0=4\pi G(\rho_{n0}+\rho_{p0})\,. 
\label{eq0-3}
\end{eqnarray}
Equations (\ref{eq0-1}) and (\ref{eq0-2}) lead to a relationship between 
chemical potentials of the neutron and the proton 
\begin{eqnarray}
\tilde\mu_{n0}-\tilde\mu_{p0}=C_{n0}-C_{p0}\,.
\end{eqnarray}
In this paper, we take $C_{n0}$ and $C_{p0}$ to be $C_{n0}=C_{p0}$, assuming 
that the two superfluids are in chemical equilibrium in the unperturbed state. 
Thus, our basic equations for obtaining unperturbed rapidly rotating stars can 
be written by 
\begin{eqnarray}
\tilde\mu_{n0}+\Phi_0-{\varpi^2\over 2}\Omega^2=C_{n0}\,,\quad 
\tilde\mu_{p0}=\tilde\mu_{n0}\,. 
\end{eqnarray}
On the other hand, the equations for the perturbed quantities are given by 
\begin{eqnarray}
\delta\tilde\mu_{n}+\delta\Phi-\varpi^2\Omega^2\delta\Omega_n=\delta C_n\,,
\label{perturbation-eq1}
\end{eqnarray}
\begin{eqnarray}
\delta\tilde\mu_{p}+\delta\Phi-\varpi^2\Omega^2\delta\Omega_p=\delta C_p \,,
\label{perturbation-eq2}
\end{eqnarray}
\begin{eqnarray}
\nabla_a\nabla^a\delta\Phi=4\pi G(\delta\rho_n+\delta\rho_p)\,.
\label{perturbation-eq3}
\end{eqnarray}
In order to solve the Poisson equations (\ref{eq0-3}) and 
(\ref{perturbation-eq3}), boundary conditions are required. The appropriate 
boundary conditions at spatial infinity and at the center of the star are 
given by 
\begin{eqnarray}
\Phi_0\rightarrow 0\,,\quad \delta\Phi\rightarrow 0\,,\quad {\rm as}\quad 
r\rightarrow\infty\,,\quad \partial_r \Phi_0=0\,,\quad \partial_r \delta\Phi=0\,,
\quad {\rm at}\quad r=0\,. 
\label{boundary-condition}
\end{eqnarray}

\subsection{Equation of state}

In order to obtain equilibrium configurations, the internal energy density 
$\cal E$ must be specified. In this paper, we employ the same analytical 
internal energy density as that used in Prix et al. (2002a), which is given by 
\begin{eqnarray}
{\cal E}={1\over 2kx_p\lbrace 1-x_p(\sigma+1)\rbrace}[x_p\rho_n^2+
\lbrace 1-x_p+\sigma(1-2x_p)\rbrace\rho_p^2-2\sigma x_p\rho_n\rho_p]\,, 
\label{def-ena}
\end{eqnarray}
where $x_p$, $\sigma$, and $k$ are constants. Note that the entrainment that 
is included in the equation of state of Prix et al. (2002a) is neglected 
because the entrainment appears in second order in $\delta\Omega_n$ or 
$\delta\Omega_p$. This internal energy density is a natural generalization of 
an $N=1$ polytropic equation of state for a barotropic ordinary fluid into the 
two superfluids because it is written in the general quadratic form of 
$\rho_n$ and $\rho_p$, given by   
\begin{eqnarray}
{\cal E}=\kappa_{nn} \rho_n^2+2\kappa_{np} \rho_n \rho_p+\kappa_{pp} 
\rho_p^2\,,
\end{eqnarray}
where $\kappa_{nn}$, $\kappa_{np}$, and $\kappa_{pp}$ are constants. 
The chemical potentials for the internal energy density 
(\ref{def-ena}) are then given by
\begin{eqnarray}
\tilde\mu_n&=&{1\over k\lbrace 1-x_p(\sigma+1)\rbrace}(\rho_n-\sigma\rho_p)
\,,\nonumber \\ 
\tilde\mu_p&=&{1\over kx_p\lbrace 1-x_p(\sigma+1)\rbrace}
[\lbrace 1-x_p+\sigma(1-2x_p)\rbrace\rho_p-\sigma x_p\rho_n]\,.  
\label{def-mu}
\end{eqnarray}
On the other hand, $\rho_n$ and $\rho_p$ can be written in terms of 
$\tilde\mu_n$ and $\tilde\mu_p$ as 
\begin{eqnarray}
\rho_n={k\over\sigma+1}[\lbrace 1-x_p+\sigma(1-2x_p)\rbrace\tilde\mu_n+
\sigma x_p\tilde\mu_p]\,,\quad  
\rho_p={kx_p\over\sigma+1}(\sigma\tilde\mu_n+\tilde\mu_p)\,, 
\end{eqnarray}
which lead to 
\begin{eqnarray}
\rho=\rho_n+\rho_p=k\lbrace (1-x_p)\tilde\mu_n+x_p\tilde\mu_p\rbrace\,. 
\end{eqnarray}
Let us consider the situation where the neutrons and the protons are in 
chemical equilibrium, which is given by the condition 
$\tilde\mu_n=\tilde\mu_p$. From equation (\ref{def-mu}), we can obtain 
\begin{eqnarray}
x_p\rho_n=(1-x_p)\rho_p\,.
\end{eqnarray}
This equation means that $\rho_n/\rho_p={\rm constant}$ inside the star and 
that $x_p$ represents the proton fraction when chemical equilibrium between 
the neutrons and the protons is achieved. On the other hand, the parameter 
$\sigma$ of the internal energy density (\ref{def-ena}) is interpreted as the 
so-called ``symmetry energy'' term (Prix et al. 2002a; Prakash, Lattimer, \& 
Ainsworth 1988). The appropriate range for $\sigma$ might be 
$\sigma\in (-1,1)$ and we consider only three values for $\sigma$, 
i.e. $\sigma=-0.5$, $0$, $0.5$, in this paper (Prix et al. 2002a).

\subsection{Unperturbed state: rapidly rotating stars}

For numerical calculations, it is convenient to introduce non-dimensional 
physical quantities as follows:
\begin{eqnarray}
&&r=r_0\hat r\,,\quad \rho_n=\rho_0\hat\rho_n\,,\quad \rho_p=\rho_0\hat\rho_p\,,\quad
\tilde\mu_n=\mu_0\hat\mu_n\,,\quad \tilde\mu_p=\mu_0\hat\mu_p\,,\quad \nonumber \\
&&\Omega_n=\sqrt{4\pi G\rho_0}\,\hat\Omega_n\,,\quad
\Omega_p=\sqrt{4\pi G\rho_0}\,\hat\Omega_p\,,\quad
\Phi=4\pi G\rho_0r_0^2\hat\Phi\,,\quad k=k_0\hat k\,,
\end{eqnarray}
where quantities with hat are non-dimensional ones, and $r_0$ and $\mu_0$ are 
defined by
\begin{eqnarray}
r_0=\sqrt{1/(4\pi Gk_0)}\,, \quad \mu_0=\rho_0/k_0\,,
\end{eqnarray}
where $\hat k$ is determined so as to be $\hat r_{\rm max}=1$ for the 
unperturbed star, and $\hat r_{\rm max}$ is the largest distance from the 
stellar surface to the center of the unperturbed star. 
Here, $r_0$ and $\rho_0$ can be given freely because of the polytropic 
equation of state.

Since, for unperturbed states, chemical equilibrium between two superfluids 
are assumed, the chemical potential $\hat\mu_{n0}$ can be written as 
$\hat\mu_{n0}=\hat k^{-1}(1-x_p)^{-1}\hat\rho_{n0}=\hat k^{-1}\hat\rho_0$ by 
virtue of the analytical equation of state (\ref{def-mu}), where $\hat\rho_0$ 
is the total mass density. Then the master equations for unperturbed stars are 
reduced to 
\begin{eqnarray}
\hat k^{-1}\hat\rho_0+\hat\Phi_0-{\hat\varpi\over 2}\,\hat\Omega^2=\hat C_{n0}\,,
\label{unp1}
\end{eqnarray}
\begin{eqnarray}
\hat\Phi_0&=&-{1\over 4\pi}\int
{\hat\rho_0(\hat{\bf r}')\over |\hat{\bf r}-\hat{\bf r}'|}d^3{\bf r}'\,,\nonumber \\
&=&-\sum_{n=0}^\infty P_{2n}(\cos\theta)\int_0^\infty\hat r'^2d\hat r' 
f_{2n}(\hat r,\hat r')\int_0^{\pi/2}\sin\theta'd\theta' P_{2n}(\cos\theta')\,
\hat\rho_0(\hat r',\theta')\,,
\label{unp2}
\end{eqnarray}
where
\begin{equation}
f_{2n}(\hat r,\hat r')=\left\{
\begin{array}{ll}
{1\over\hat r}\left({\hat r'\over\hat r}\right)^{2n} & {\rm for}\quad \hat r' < \hat r\,, \\
{1\over\hat r'}\left({\hat r\over\hat r'}\right)^{2n} & {\rm for}\quad \hat r' \ge \hat r\,, 
\end{array}
\right.
\end{equation} 
and $P_{2n}(\cos\theta)$ are the Legendre polynomials. Here, the gravitational 
potential $\hat\Phi_0$ has been written in the integral representation, in 
which a proper Green function is employed to include physically appropriate 
boundary conditions both at $\hat r=0$ and at spatial infinity 
(\ref{boundary-condition}). In this integral representation, equatorial 
symmetry of the matter distribution has been assumed because we are concerned 
about stars having this equatorial symmetry. Note that these equations are 
exactly the same as those of polytropic ordinary fluid stars with polytrope 
index $N=1$. Thus, solutions to equations (\ref{unp1}) and (\ref{unp2}) can be 
obtained by the so-called Hachisu's Self Consistent Field scheme (HSCF), once 
the axis ratio, $r_p$, is given (Hachisu 1986). Here, $r_p$ is defined by
\begin{equation}
r_p=\hat r_{\rm min}/\hat r_{\rm max}=\hat r_{\rm min}\,,
\end{equation}
where $\hat r_{\rm min}$ is the minimum distance from the stellar surface to 
the center of the star. Note that solutions $\hat\rho_0$, $\hat\Phi_0$, and 
$\hat C_{n0}$ are universal functions in the sense that those are independent 
of the proton fraction $x_p$. After getting a solution $\hat\rho_0$, and 
giving the proton fraction $x_p$, we can obtain the mass densities and 
chemical potentials of two superfluids through the formulas 
\begin{eqnarray}
\hat\rho_{n0}=(1-x_p)\hat\rho_0\,,\quad \hat\rho_{p0}=x_p\hat\rho_0\,,\quad 
\hat\mu_{n0}=\hat\mu_{p0}=\hat\rho_0\hat k^{-1}\,. 
\end{eqnarray}

The surface of the star $R(\theta)$ for unperturbed stars is defined by    
\begin{eqnarray}
P(R(\theta),\theta)=0\,,
\label{def-surface}
\end{eqnarray}
where $P$ is the generalized pressure, which is given, for our internal energy 
density, by 
\begin{eqnarray}
P={\mu_0\rho_0\over 2\hat k x_p\lbrace 1-x_p(\sigma+1)\rbrace}
[x_p\hat\rho_n^2+\lbrace 1-x_p+\sigma(1-2x_p)\rbrace\hat\rho_p^2-
2\sigma x_p\hat\rho_n\hat\rho_p]\,\,.
\end{eqnarray}
Pressure $P_0$ for the unperturbed state can be explicitly written as  
\begin{eqnarray}
P_0={\mu_0\rho_0\over 2\hat k(1-x_p)^2}\hat\rho_{n0}^2=
{\mu_0\rho_0\over 2\hat k}\hat\rho_0^2\,. 
\end{eqnarray}
The non-dimensional stellar surface $\hat R_0(\theta)$ for an unperturbed star 
is then given by a solution of the algebraic equation 
$\hat\rho_0(\hat R_0(\theta),\theta)=0$.

\subsection{Effect of the rotation velocity difference on the stellar structure}

In this paper, we consider perturbed states whose densities of two superfluids 
$\rho_n$ and $\rho_p$ at the center of the star have the same values as those 
of unperturbed states. Because of our equations of state (\ref{def-mu}), this 
requirement is equivalent to the assumptions of $\delta\mu_n=\delta\mu_p=0$ at 
the center of the star. Thus, the integral constants in equations 
(\ref{perturbation-eq1}) and (\ref{perturbation-eq2}) may be determined so as 
to be  
\begin{equation}
\delta C_n=\delta C_p=\delta\Phi(r=0,\theta)\,. 
\label{perturbed_constants}
\end{equation}
For the analytical internal energy density (\ref{def-ena}), perturbations of 
the total mass density $\delta\hat\rho$ is given in terms of perturbed 
chemical potentials $\delta\hat\mu_n$ and $\delta\hat\mu_p$ as 
\begin{eqnarray}
\delta\hat\rho=\delta\hat\rho_n+\delta\hat\rho_p=
{\hat k} \left\{(1-x_p)\delta\hat\mu_n+x_p\delta\hat\mu_p \right\} \,.  
\end{eqnarray}
From equations (\ref{perturbation-eq1}) and (\ref{perturbation-eq2}), thus, 
we obtain 
\begin{eqnarray}
\hat k^{-1} \delta\hat\rho+\delta\hat\Phi-\hat\varpi^2\hat\Omega^2
\lbrace(1-x_p)\delta\Omega_n+x_p\delta\Omega_p \rbrace=\delta\hat C_n\,, 
\label{peq-1}
\end{eqnarray}
where the relations of the integral constants (\ref{perturbed_constants}) have 
been assumed. Note that the perturbed total density is dependent on the proton 
fraction $x_p$ but not on the symmetry energy parameter $\sigma$. 
From equation (\ref{peq-1}), it is found that perturbations $\delta\hat\rho$, 
$\delta\hat\Phi$, and $\delta\hat C_n$ can be represented in terms of three 
functions independent of $\delta\Omega_n$, $\delta\Omega_p$, and $x_p$ as 
follows: 
\begin{eqnarray}
\delta\hat\rho&=&\lbrace(1-x_p)\delta\Omega_n+x_p\delta\Omega_p \rbrace\delta\bar\rho\,,\quad 
\delta\hat\Phi=\lbrace(1-x_p)\delta\Omega_n+x_p\delta\Omega_p \rbrace\delta\bar\Phi\,,\nonumber \\
\delta\hat C_n&=&\lbrace(1-x_p)\delta\Omega_n+x_p\delta\Omega_p \rbrace\delta\bar C_n\,, 
\end{eqnarray}
where the three functions $\delta\bar\rho$, $\delta\bar\Phi$, and 
$\delta\bar C_n$ are the solutions of the equations 
\begin{eqnarray}
\hat k^{-1} \delta\bar\rho+\delta\bar\Phi-\hat\varpi^2\hat\Omega^2=\delta\bar C_n\,, 
\label{peq1}
\end{eqnarray}
\begin{eqnarray}
\hat\nabla_a\hat\nabla^a\delta\bar\Phi=\delta\bar\rho\,. 
\label{peq2}
\end{eqnarray}
Note that three quantities $\delta\bar\rho$, $\delta\bar\Phi$, and 
$\delta\bar C_n$ are universal functions in the sense that those are 
independent of the parameters $x_p$ and $\sigma$, and that those are dependent 
only on the structure of the unperturbed star.  With equations 
(\ref{perturbation-eq1}) and (\ref{perturbation-eq2}), the chemical potentials 
of two superfluids are given by 
\begin{eqnarray}
\delta\hat\mu_n&=&
\lbrace x_p(\delta\bar\Phi-\delta\bar C_n)+\hat k^{-1}\delta\bar\rho\rbrace\delta\Omega_n-
x_p(\delta\bar\Phi-\delta\bar C_n)\delta\Omega_p\,,\nonumber\\
\delta\hat\mu_p&=&
(1-x_p)(\delta\bar C_n-\delta\bar\Phi)\delta\Omega_n+  
\lbrace (1-x_p)(\delta\bar\Phi-\delta\bar C_n)+\hat k^{-1}\delta\bar\rho\rbrace\delta\Omega_p
\,. 
\end{eqnarray}
Perturbations of the mass densities, on the other hand, can be written in 
terms of $\delta\hat\mu_n$ and $\delta\hat\mu_p$ as 
\begin{eqnarray}
\delta\hat\rho_n={\hat k\over\sigma+1}\left[\lbrace 
1-x_p+\sigma(1-2x_p)\rbrace\delta\hat\mu_n+ \sigma x_p\delta\hat\mu_p\right]\,,\quad
\delta\hat\rho_p={\hat k x_p\over\sigma+1}(\sigma\delta\hat\mu_n+\delta\hat\mu_p)\,,
\end{eqnarray}
Note that, for our internal energy density, the chemical potentials do not 
depend on the symmetry energy parameter $\sigma$, while the mass densities do. 
The perturbations of the generalized pressure $\delta P$ is given by
\begin{eqnarray}
\delta P=\mu_0\rho_0\hat k^{-1}\hat\rho_0\{(1-x_p)\delta\hat\mu_n+x_p\delta\hat\mu_p\}\,. 
\end{eqnarray}
If we write the stellar surface with accuracy up to the first order of the 
rotation velocity difference as 
\begin{eqnarray}
\hat R(\theta)=\hat R_0(\theta)+\delta\hat R(\theta)\,,  
\end{eqnarray}
the first order corrections for the stellar surface $\delta\hat R(\theta)$ can 
be due to equation (\ref{def-surface}) written in terms of $\delta\hat\mu_n$ 
and $\delta\hat\mu_p$ as 
\begin{eqnarray}
\delta\hat R(\theta)=(1-x_p)\delta\hat R_n+x_p\delta\hat R_p\,,
\end{eqnarray}
where $\delta\hat R_n$ and $\delta\hat R_n$ are defined by 
\begin{eqnarray}
\delta\hat R_n=-{\delta\hat\mu_n\over\partial_r\hat\rho_0}
(\hat r=\hat R_0(\theta),\theta)\,,\quad 
\delta\hat R_p=-{\delta\hat\mu_p\over\partial_r\hat\rho_0}
(\hat r=\hat R_0(\theta),\theta)\,. 
\end{eqnarray}
Because $\hat\mu_n(\hat R_0+\delta\hat R_n)=0$ and 
$\hat\mu_p(\hat R_0+\delta\hat R_p)=0$ are satisfied, two surfaces of 
$\hat r=\hat R_0+\delta\hat R_n$ 
and $\hat r=\hat R_0+\delta\hat R_p$ can be interpreted as the surfaces of the 
neutron and proton superfluids, respectively. Although we can define the 
respective fluid surfaces of two superfluids as the zero-density surfaces 
(Prix 1999; Prix et al. 2002a), we make use of the definition with the 
chemical potential for the surfaces because the definition with the chemical 
potentials is considered to be natural generalization of the surface 
definition for ordinary fluid stars, for which the surface is defined as a 
zero-pressure surface not as a zero-density surface. 
Note that as we can see from equation (35), in general, ``zero-chemical 
potential surfaces'' do not coincide with ``zero-density surfaces''. 
In other words, the densities do not necessarily vanish on the 
zero-chemical potential surfaces because the equi-potential surfaces in 
general incline to the equi-density surfaces in the two-fluid model.

In our numerical procedure of solving equations (\ref{peq1}) and (\ref{peq2}), 
in order for the Poisson equation (\ref{peq2}) to satisfy the boundary 
condition (\ref{boundary-condition}) explicitly, equation (\ref{peq2}) is 
converted into the integral representation, given by
\begin{eqnarray}
\delta\bar\Phi=-\sum_{n=0}^\infty P_{2n}(\cos\theta)
\int_0^{\pi/2}\sin\theta'd\theta'\int_0^{\hat R_0(\theta')}\hat r'^2d\hat r' 
f_{2n}(\hat r,\hat r')P_{2n}(\cos\theta')\,\delta\bar\rho(\hat r',\theta')\,. 
\label{p2}
\end{eqnarray}
In this paper, equations (\ref{peq1}) and (\ref{p2}) are numerically solved 
with a variant of the so-called Self Consistent Field scheme (Ostriker \& Mark 
1968). To obtain solutions, we follow the following steps: i) By assuming an 
initial guess for $\delta\bar\rho$, compute the gravitational potential 
through the two-dimensional integration (\ref{p2}). ii) Determine the value of 
parameter $\delta\bar C_n$ from equation (\ref{perturbed_constants}). iii) By 
using the obtained $\delta\bar\Phi$ and $\delta\bar C_n$, solve equation 
(\ref{peq1}) for the perturbed density $\delta\bar\rho$. iv) Compare the 
obtained perturbed density with the one used in obtaining the perturbed 
potential. If the relative changes of $\delta\bar\rho$ 
are less than $10^{-8}$ at all grid points, then the obtained perturbed 
density distribution is considered to be a converged solution. If the 
condition for the relative changes is not satisfied, go back to step ii) and 
there the obtained perturbed density is treated as a new initial guess for 
$\delta\bar\rho$ for the next iteration cycle.

In actual computations, we make use of equidistantly spaced discrete meshes in 
the radial direction ($0\le\hat r\le \hat r_{\rm max}=1$). In order to 
calculate integrations in $r$ direction, we employ a classical trapezoidal 
rule. As for the angular variable $\theta$, we take the angular grid points 
at $\mu_i=\cos\theta_i$, where $\mu_i$'s are zeros of the Legendre polynomial 
of order $2L-1$, i.e. $P_{2L-1}(\mu_i)=0$, and there are $L$ grid points for 
$0\le\theta\le\pi/2$. We employ the Gaussian quadrature (see, e.g., Abramowitz 
\& Stegun 1964) to evaluate integrations in the $\theta$ direction. In the 
present investigation, we take the number of mesh points to be $500\times 25$ 
$(r\times\theta)$. The Legendre polynomials in equations (\ref{unp2}) and 
(\ref{p2}) are added up to $P_{46}(\mu)$.

\section{Numerical Results}

In this paper, we assume the proton fraction to be $x_p=0.1$ because a typical 
proton fraction in the cores of old neutron stars is expected to be around 
$x_p=0.1$. For the parameter of the symmetry energy, $\sigma$, since we have 
little information about the range of $\sigma$, we investigate three cases,  
i.e. $\sigma=-0.5\,,\ 0\,,\ 0.5$, in order to examine the effect of the 
symmetry energy on the stellar structures. Note that the total density of 
superfluids and the chemical potentials are dependent on the proton fraction 
but not on the symmetry energy. We exhibit the results for two cases of 
$(\delta\Omega_n,\delta\Omega_p)=(1,0)$ and 
$(\delta\Omega_n,\delta\Omega_p)=(0,1)$ because solutions for other values of 
$(\delta\Omega_n,\delta\Omega_p)$ are expressed in terms of linear 
superpositions of those two solutions.

First, let us consider the unperturbed stars. In our unperturbed states, two 
superfluids are assumed to be in the same rotational motion and in chemical 
equilibrium. As mentioned in the last section, the structures of unperturbed 
stars are, therefore, almost the same as those of uniformly rotating $N=1$ 
polytropic stars. Thus, there are no new features for the unperturbed stars. 
For reader's convenience, however, we exhibit some results for the unperturbed 
stars. In Figures 1 through 4, we show the non-dimensional fundamental 
quantities of the unperturbed stars, the axis ratios, $r_p$, the total masses, 
$\hat M$, the total moments of inertia, $\hat I$, and the ratios of the 
rotational energy to the absolute value of the gravitational energy, $T/|W|$,
as functions of the angular velocity, $\bar\Omega$, where $\bar\Omega$ is the 
non-dimensional angular velocity of the star, defined by 
$\bar\Omega=\Omega/(GM/R^3)^{1/2}$, and $R$ denotes the stellar radius on 
the equatorial plane. Here, the total mass, $M$ is defined by
\begin{equation}
M=\int\rho_0d^3{\bf r}={4\pi \over 3}\rho_0r_0^3\hat M=M_0\hat M\,.
\end{equation}
The total moments of inertia, $I$, is given by 
\begin{equation}
I=\int\rho_0\varpi^2d^3{\bf r}=4\pi\rho_0r_0^5\hat I=I_0\hat I\,.
\end{equation}
The rotational energy, $T$, and the gravitational energy, $W$, are, 
respectively, defined by 
\begin{equation}
T={1\over 2}\int\rho_0\varpi^2\Omega^2d^3{\bf r}\,,
\quad 
W=-{1\over 2}\int\rho_0\Phi_0d^3{\bf r}\,. 
\end{equation}
Since $\rho_n/\rho_p={\rm const}$ for the unperturbed stars, we can write the 
masses, $M_n$, $M_p$, and the moments of inertia, $I_n$, $I_p$, for the 
neutrons and the protons as 
\begin{equation}
M_n=(1-x_p)M\,,\quad M_p=x_p M\,,\quad I_n=(1-x_p)I\,,\quad I_p=x_pI\,. 
\end{equation}

In order to check our numerical code, we have calculated a solution for a
slowly rotating star with the axis ratio $r_p=0.992$, whose angular velocity 
is $\bar\Omega=0.10298$, and compared them with the analytical solution 
obtained by Prix et al. (2002a), in which the structure of a slowly 
rotating superfluid star was examined analytically with the slow rotation 
approximation. The perturbed densities obtained from two methods are shown as 
functions of $\hat r$ in Figure 5. The solid lines show the analytical result 
for $\delta\hat\rho_n$ with the slow rotation approximation, and the solid 
squares our numerical result without assuming the slow rotation approximation. 
The parameters of the model shown in Figure 5 have been chosen to be 
$\sigma=-0.5$ and $(\delta\Omega_n,\delta\Omega_p)=(1,0)$. As seen from this 
figure, our result is in good agreement with the analytical result of Prix 
et al. (2002a) as long as the angular velocity of the star is small enough.

In Figures 6 through 11, typical density distributions for the neutrons and 
the protons, $\delta\hat\rho_n$ and $\delta\hat\rho_p$ are represented for 
three values of the symmetry energy parameter, $\sigma=-0.5\,,\ 0\,,\ 0.5$. 
For the models shown in Figures 6 through 11, the axis ratio $r_p$ for the 
unperturbed star has been taken to be $r_p=0.668$, and the corresponding 
angular velocity is given by $\bar\Omega=0.8037$. Figures 6 through 8 show the 
density distributions for $(\delta\Omega_n,\delta\Omega_p)=(1,0)$, and 
Figures 9 through 11 for $(\delta\Omega_n,\delta\Omega_p)=(0,1)$. In those 
figures, the perturbed densities are shown versus the radial coordinate 
$\hat r$ for six different values of $\theta_i$, in which the longest curve 
corresponds to the result on the equator, $\theta=\pi/2$, and $\theta_i$'s 
decrease for each successively shorter curve toward the value $\theta=0$ on 
the symmetry axis. Comparing Figure 6 (left panel) with Figure 5, in which 
distributions of $\delta\hat\rho_n$ are shown for the same parameter as those 
in Figure 6 except for the axis ratio $r_p$ or for the rotation velocity 
$\bar\Omega$, we observe that the basic qualitative properties of the density 
perturbations for two superfluids do not depend on the value of $\bar\Omega$ 
so much. We do not therefore show the density distributions for other rotation 
rates $\bar\Omega$ in this paper. From Figures 6 through 11, we can see that 
the amplitudes of $\delta\hat\rho_n$ are much larger than those of 
$\delta\hat\rho_p$ for the solutions of 
$(\delta\Omega_n,\delta\Omega_p)=(1,0)$, while the amplitudes of 
$\delta\hat\rho_n$ is smaller than those of $\delta\hat\rho_p$ for the case of 
$(\delta\Omega_n,\delta\Omega_p)=(0,1)$. As shown in Figures 9 through 11, 
however, for the solutions of $(\delta\Omega_n,\delta\Omega_p)=(0,1)$, 
difference of the amplitude between $\delta\hat\rho_n$ and $\delta\hat\rho_p$ 
is not so large. This is because the proton fraction is not very large in the 
interior. Similar behaviors were found by Prix et al. (2002a). It is also 
found from Figures 6 through 11 that $\delta\hat\rho_p$ is strongly dependent 
on the values of $\sigma$ for the $(\delta\Omega_n,\delta\Omega_p)=(1,0)$ 
case, while $\delta\hat\rho_n$ depends strongly on $\sigma$ for the 
$(\delta\Omega_n,\delta\Omega_p)=(0,1)$ case. This means that the significant 
parameters to model a superfluid neutron star are appropriate combinations
of the symmetry energy parameter $\sigma$ and the velocity difference between
two superfluids.

In Figures 12 and 13, the perturbed axis ratios are plotted as 
functions of the angular velocity $\bar\Omega$. Here, the functions, 
$\delta r_p$, $\delta r_{p,n}$, and $\delta r_{p,p}$ can be interpreted as 
the perturbations of the axis ratios of the whole star, the neutron superfluid,
and the proton superfluid, respectively, and are defined by  
\begin{eqnarray}
\delta r_p&=&\delta\hat R(\theta=0)-r_p\delta\hat R(\theta=\pi/2)\,,\quad
\delta r_{p,n}=\delta\hat R_n(\theta=0)-r_p\delta\hat R_n(\theta=\pi/2)
\,,\nonumber \\
\delta r_{p,p}&=&\delta\hat R_p(\theta=0)-r_p\delta\hat R_p(\theta=\pi/2)\,. 
\end{eqnarray}
The results for $(\delta\Omega_n,\delta\Omega_p)=(1,0)$ and 
$(\delta\Omega_n,\delta\Omega_p)=(0,1)$ are shown in Figures 12 and 13, 
respectively. Note that in these figures, the perturbed axis ratios 
are shown for $0\le\bar\Omega\le 0.9$ because they diverge as $\bar\Omega$ 
goes to its maximum value. These figures illustrate how the surface of the 
star is deform due to the effect of the rotation velocity difference between 
two superfluids. It is found that the change of the axis ratio of the rotating 
component with the slower angular velocity becomes positive when the angular 
velocity $\bar\Omega$ becomes larger than some critical value. In other words, 
the rotating component with the slower angular velocity tends to become 
prolate when a star rotates very rapidly.

In Figure 14, the perturbed mass is shown as a function of 
$\bar\Omega$. Here, the perturbed total mass $\delta M$ is defined as 
\begin{equation}
\delta M={3\over 4\pi \hat M}\int\delta\hat\rho d^3{\bf r}\,.
\end{equation}
The solid curve and the dashed curve show $\delta M$ for the solutions with 
$(\delta\Omega_n,\delta\Omega_p)=(1,0)$ and 
$(\delta\Omega_n,\delta\Omega_p)=(0,1)$, respectively. In this paper, we 
consider the perturbed states whose central density is the same as that of 
the unperturbed states. Thus, the mass of the perturbed stars can change. 
In Figures 15 and 16, we plot the perturbed moments of inertia for the neutron 
and the proton as functions of the angular velocity $\bar\Omega$. Here, the 
perturbed moments of inertia for the neutron and the proton, $\delta I_n$ and 
$\delta I_p$, are defined as 
\begin{equation}
\delta I_n={1 \over 4\pi r_0^5(1-x_p)\hat I}\int\delta\hat\rho_n\varpi^2d^3{\bf r}\,,
\quad 
\delta I_p={1 \over 4\pi r_0^5x_p\hat I}\int\delta\hat\rho_p\varpi^2d^3{\bf r}\,. 
\end{equation}
Figures 15 and 16 show the perturbations of the moments of inertia for 
$(\delta\Omega_n,\delta\Omega_p)=(1,0)$ and 
$(\delta\Omega_n,\delta\Omega_p)=(0,1)$, respectively. In these figures, 
results for three different values of $\sigma$, $\sigma=-0.5,\ 0,\ 0.5$ are 
displayed. Figures 15 and 16 show that the perturbed moments of inertia for 
the proton are strongly dependent on the symmetry energy parameter 
$\sigma$ for $(\delta\Omega_n,\delta\Omega_p)=(1,0)$ case, while the perturbed
moments of inertia for the neutrons strongly depend on $\sigma$ for 
$(\delta\Omega_n,\delta\Omega_p)=(0,1)$ case. This is consistent with that of 
the properties of the density distributions for the neutrons and the protons.  
It is noted that amplitudes of $\delta I_n$ and $\delta I_p$ become large as 
the angular velocity $\bar\Omega$ is increased because the perturbations are 
roughly proportional to $\hat\Omega^2$ due to equation (\ref{peq1}).

\section{Summary and discussion}

In this paper we have developed a formulation for constructing and examining 
rapidly rotating neutron stars that contain two superfluids, taking account of 
the effect of the rotation velocity difference between two superfluids. We 
assumed neutron stars to be composed of the superfluid neutrons and the 
mixture of the superfluid protons and the normal fluid electrons. To describe 
Newtonian dynamics of the two superfluids, the Newtonian version of the 
two-fluid formalism developed by Prix (2002) was employed. In this paper, we 
considered the situation where two superfluids rapidly rotate around the same 
rotation axis, but the rotation velocity difference between two superfluids 
is very small. We then treated the effect of the rotation velocity difference 
on the equilibrium configurations as a small perturbation to rapidly rotating 
superfluid stars whose rotation velocities of two superfluids are the same. 
We derived basic equations for perturbations on the structure of rapidly 
rotating superfluid stars due to the rotation velocity difference between two 
superfluids. Assuming the superfluids to obey a simple analytical equation of 
state used by Prix et al. (2002a), we obtained numerical solutions for the 
perturbed quantities and found that the density distributions of the 
superfluids are strongly dependent of the symmetry energy parameter $\sigma$, 
which appears in the analytical equation of state. Similar properties were 
found in Prix et al. (2002a). It was also found that if Prix et al. (2002a)'s 
analytical 
equation of state is assumed, the perturbations can be represented in terms of 
the universal functions that are independent of the parameters of the equation 
of state.

Although we only considered one special equation of state for superfluids in 
the present investigation, the formalism we derived in this paper is 
straightforwardly applicable to general equations of state. We treated the 
superfluids inside neutron stars in the framework of Newtonian dynamics. The 
effect of general relativity must not be however neglected for the structures 
of neutron stars because neutron stars are quite compact in a sense that 
a general relativistic effect can be expressed by the factor $GM/c^2R$ whose 
typical value is $\sim 0.2$ for neutron stars, where $c$ is the speed of 
light. It is straightforward to extend the present formulation to general 
relativistic configurations and we will do it in the future. Recently, Prix 
et al. (2002b) have obtained rapidly rotating relativistic superfluid stars 
whose rotation velocity of the neutrons differs from that of the protons. 
They did not assume that the rotation velocity difference between two fluids 
is very small. In other words, their method can be applied to equilibrium 
configurations with any rotation velocities. Yet, our perturbation method 
developed in this investigation has an advantage in studying structures of 
real neutron stars because it would give results with higher accuracy by 
taking account of the smallness of the velocity difference directly, as long 
as the rotation velocity difference between two fluids in real neutron stars 
is very small. 
A solid crust is believed to exist near the surface of cold and old neutron 
stars, and can have a significant influence on the structure of the 
equilibrium states if the solid crust is not in a strain-free state in the 
equilibrium states of rotating neutron stars. Therefore, an investigation of 
the effect of the solid crust on the equilibrium states remains as 
a challenging problem in the future too.

\vskip 0.5cm
\noindent{\sl  Acknowledgments:} 
{S.Y. acknowledges financial support from Funda\c c\~ao para a  Ci\^encia e a Tecnologia 
(FCT) through project SAPIENS 36280/99.}

\newpage 

\begin{figure}
\centering
\includegraphics[width=8cm]{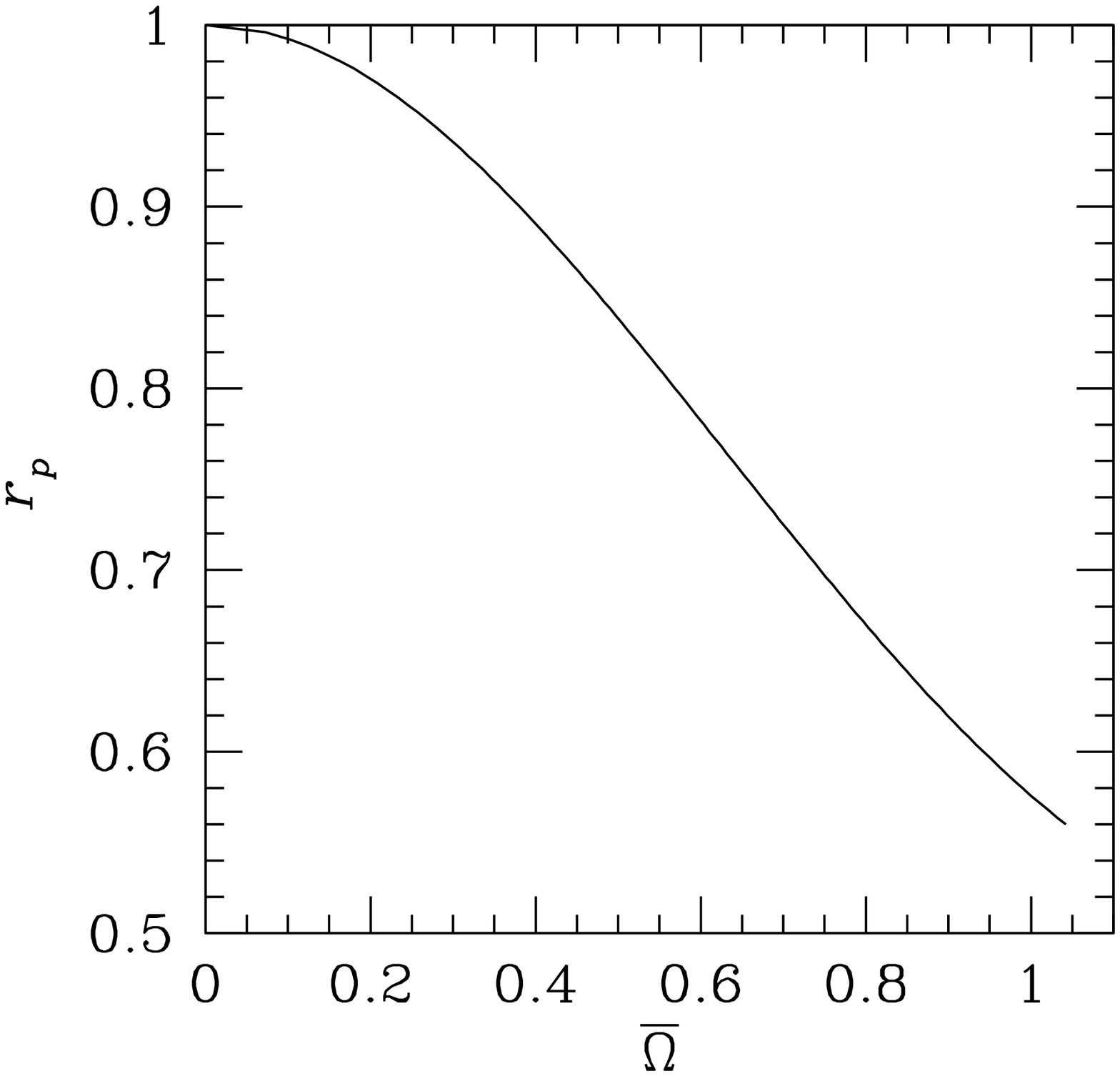}
\caption{Axis ratio $r_p$ for the unperturbed stars is given as a function of 
$\bar\Omega$}
\end{figure}
\begin{figure}
\centering
\includegraphics[width=8cm]{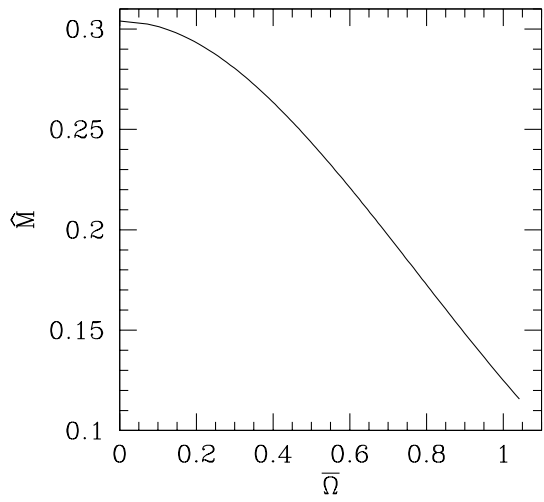}
\caption{Same as Figure 1 but for the mass $\hat{M}$}
\end{figure}
\begin{figure}
\centering
\includegraphics[width=8cm]{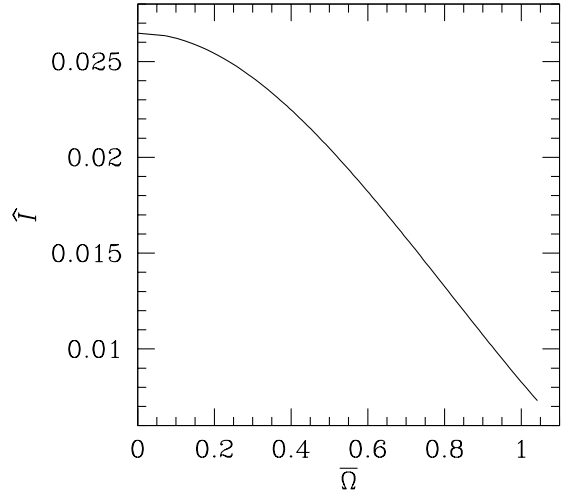}
\caption{Same as Figure 1 but for the moment of inertia $\hat{I}$}
\end{figure}
\begin{figure}
\centering
\includegraphics[width=8cm]{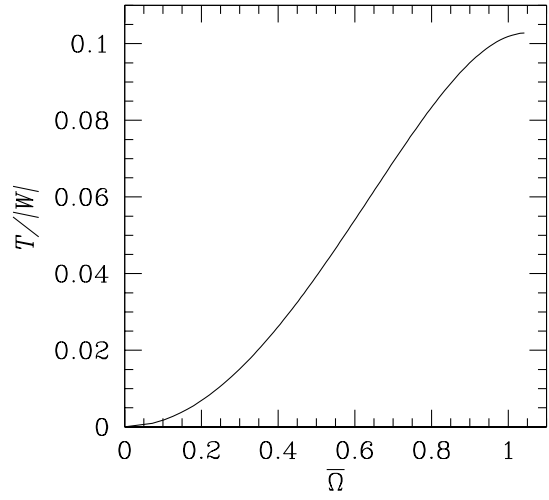}
\caption{Same as Figure 1 but for the ratios of the rotational energy to 
the absolute value of the gravitational energy $T/|W|$}
\end{figure}
\begin{figure}
\centering
\includegraphics[width=8cm]{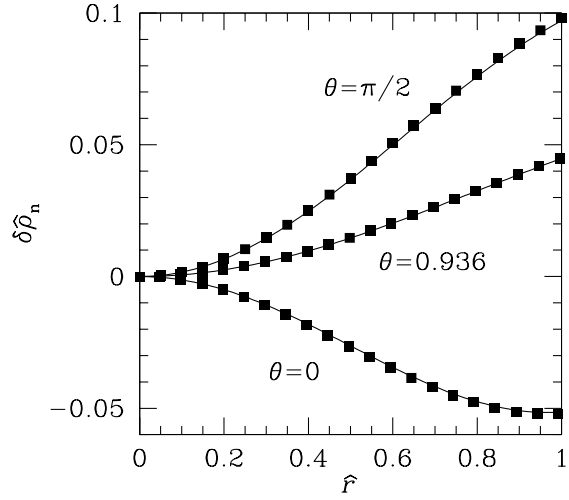}
\caption{Perturbed densities for the neutron $\delta\hat\rho_n$ are
given as functions of $\hat r$. The solid curves and solid squares are used 
to indicate $\delta\hat\rho_n$ obtained from Prix et al. (2002a)'s analytical 
formula and our numerical scheme, respectively. The model parameters
are $\bar\Omega=0.10298$, $\sigma=-0.5$, 
and $(\delta\Omega_n,\delta\Omega_p)=(1,0)$.}  
\end{figure}
\begin{figure}
\centering
\includegraphics[width=8cm]{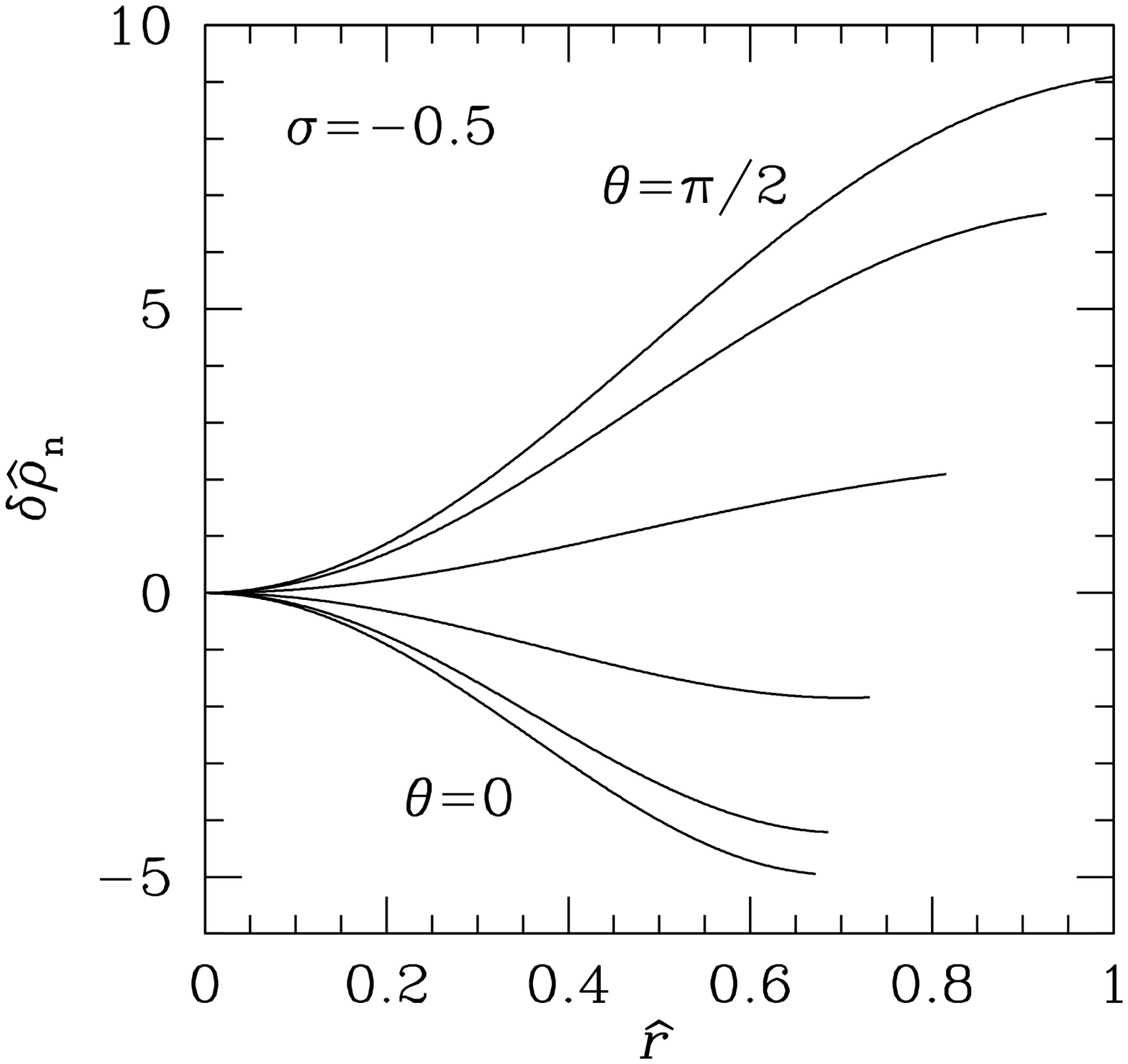}
\includegraphics[width=8cm]{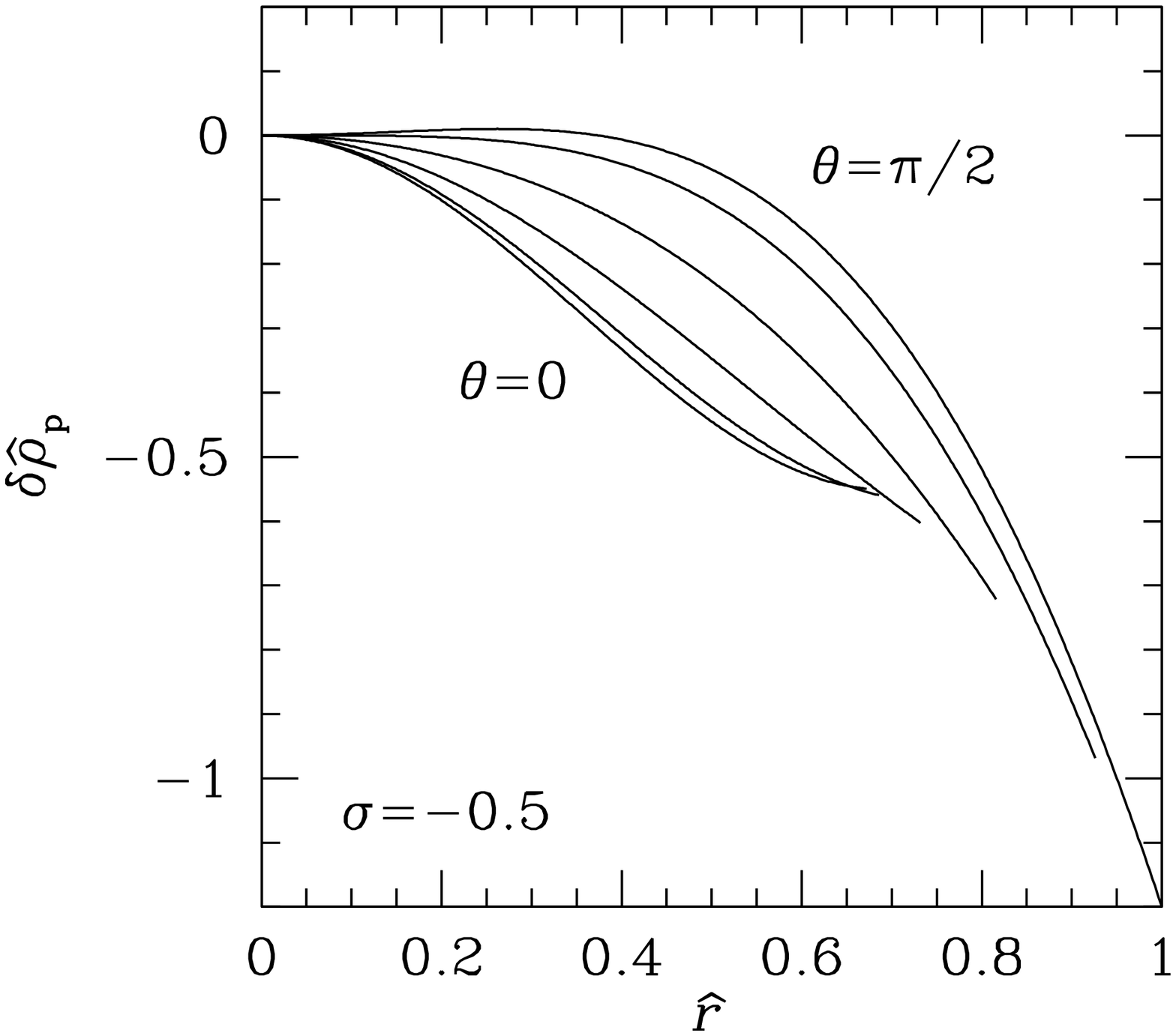}
\caption{Perturbed densities for the neutron (left) and the proton (right) 
are given as functions of $\hat r$ for six different values of $\theta$. 
The model parameters are $\bar\Omega=0.8032$, $\sigma=-0.5$, and 
$(\delta\Omega_n,\delta\Omega_p)=(1,0)$.}
\end{figure}
\begin{figure}
\centering
\includegraphics[width=8cm]{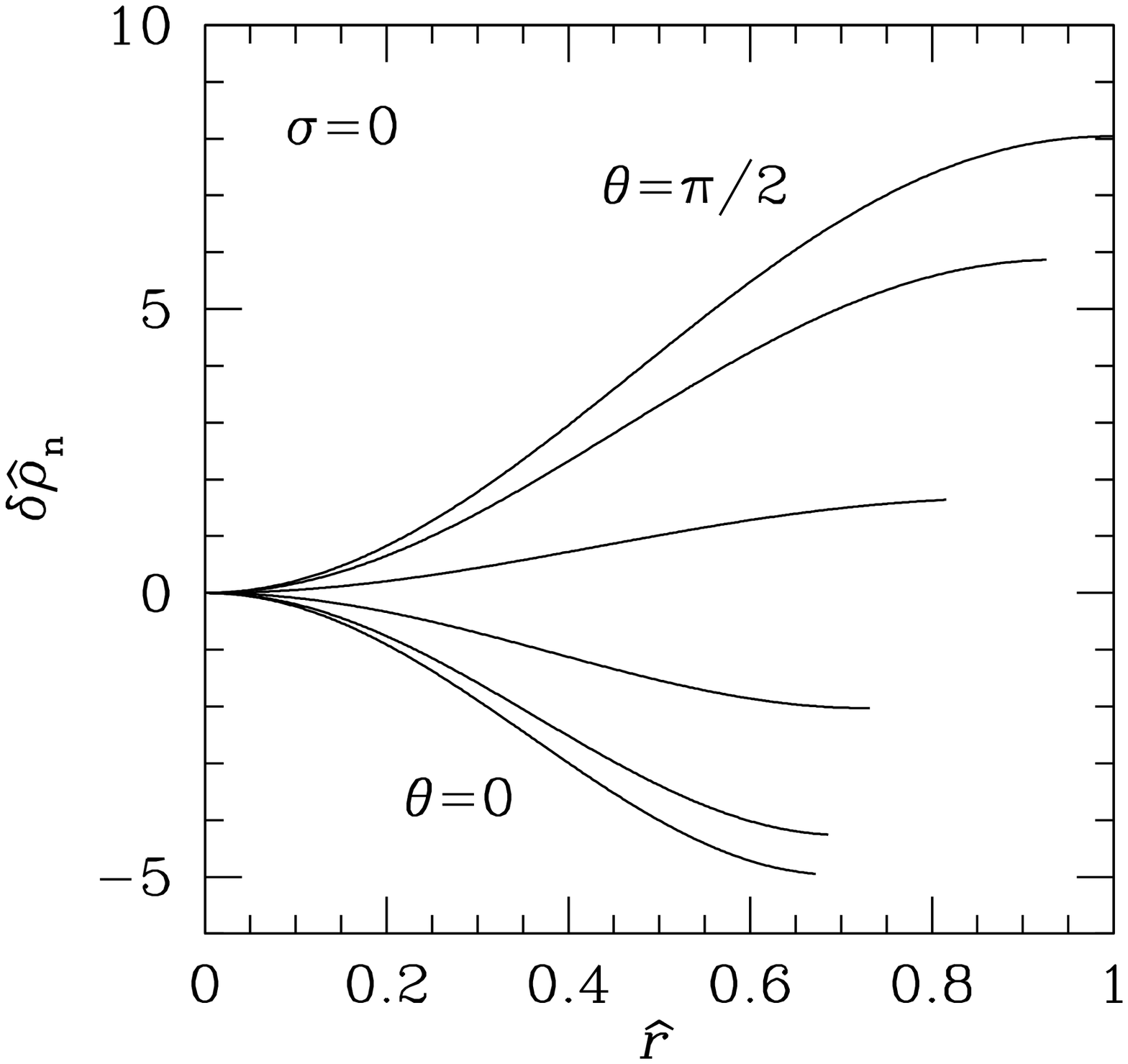}
\includegraphics[width=8cm]{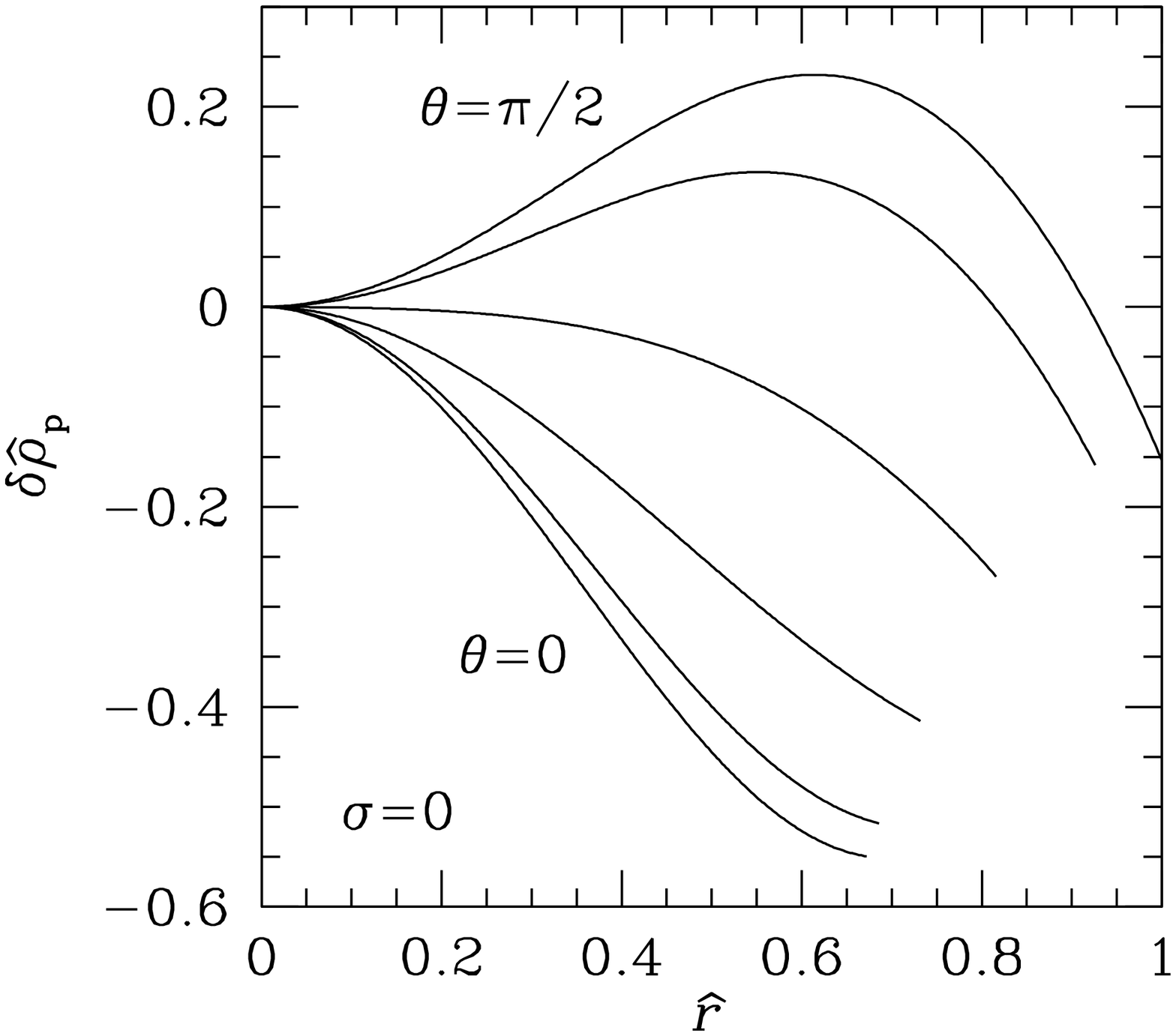}
\caption{Same as Figure 6 but for $\sigma=0$.} 
\end{figure}
\begin{figure}
\centering
\includegraphics[width=8cm]{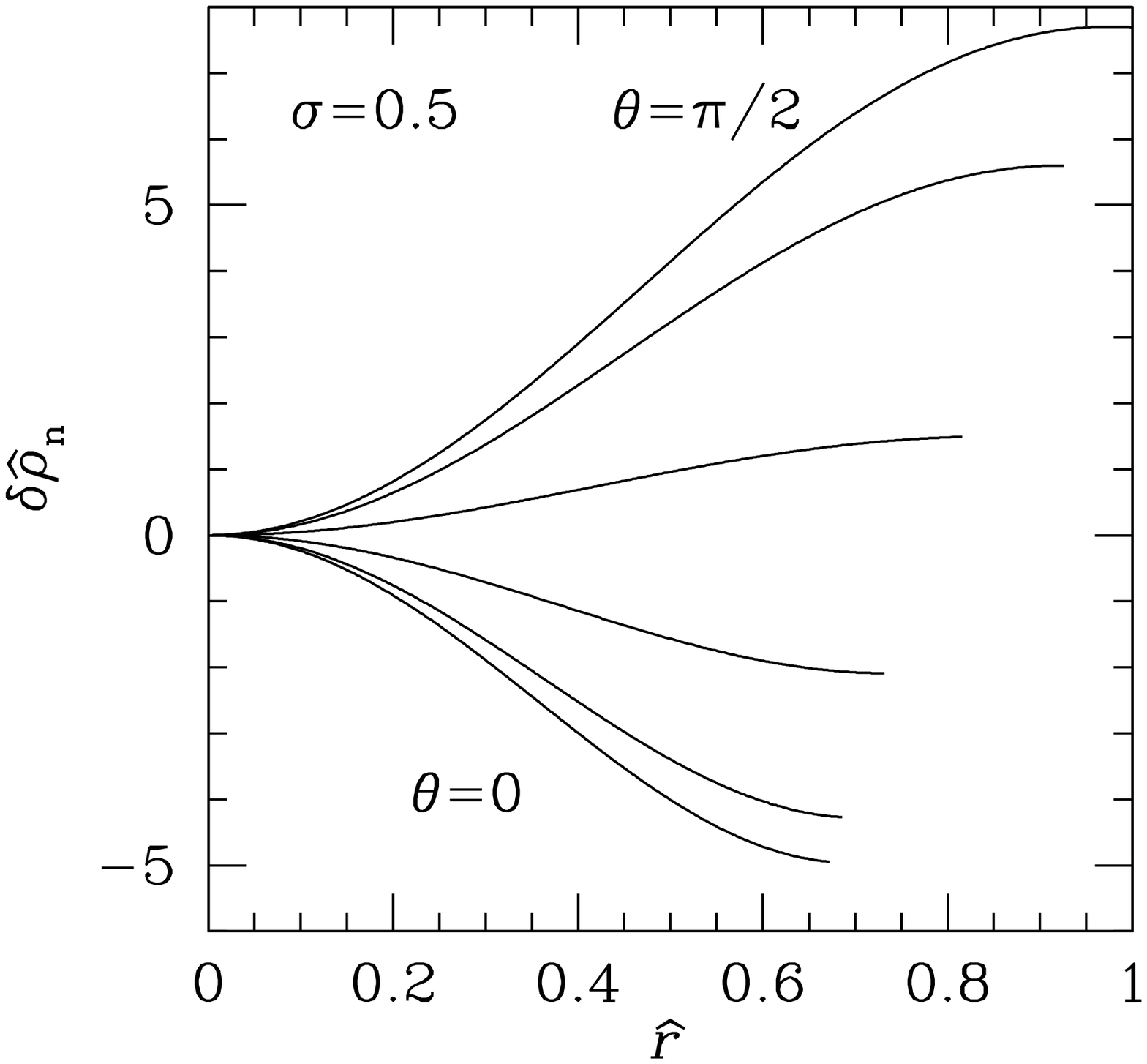}
\includegraphics[width=8cm]{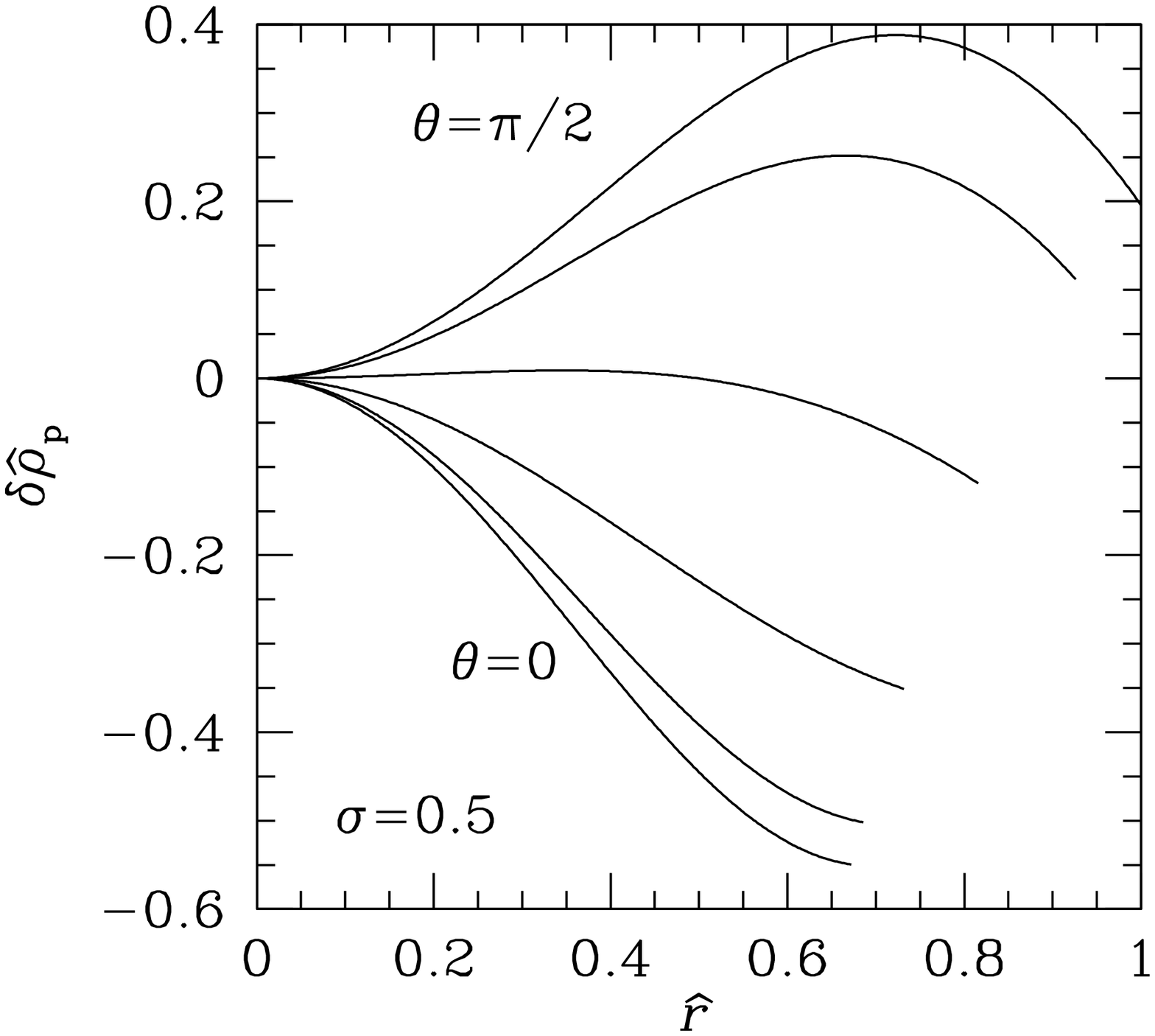}
\caption{Same as Figure 7 but for $\sigma=0.5$.}
\end{figure}
\begin{figure}
\centering
\includegraphics[width=8cm]{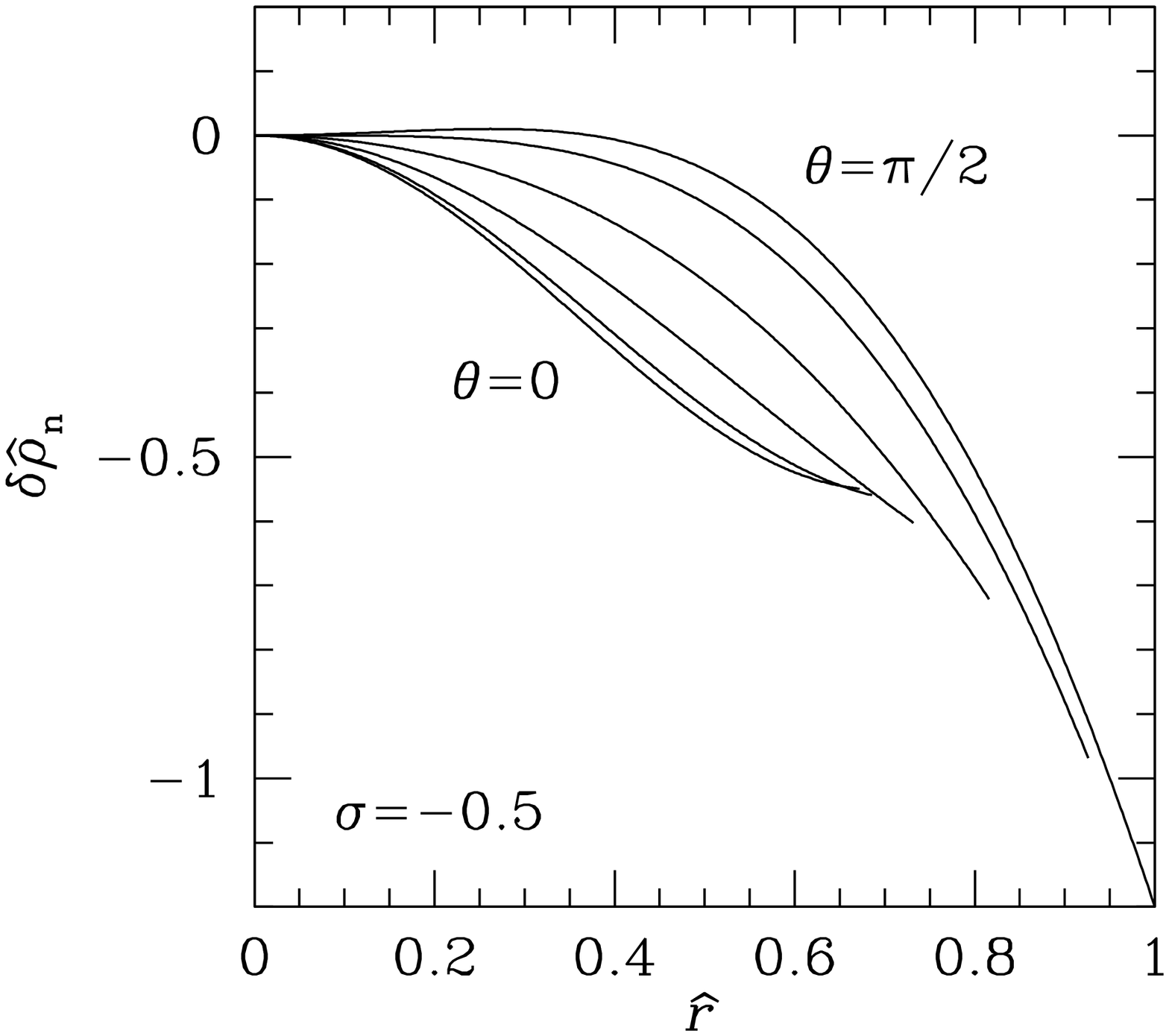}
\includegraphics[width=8cm]{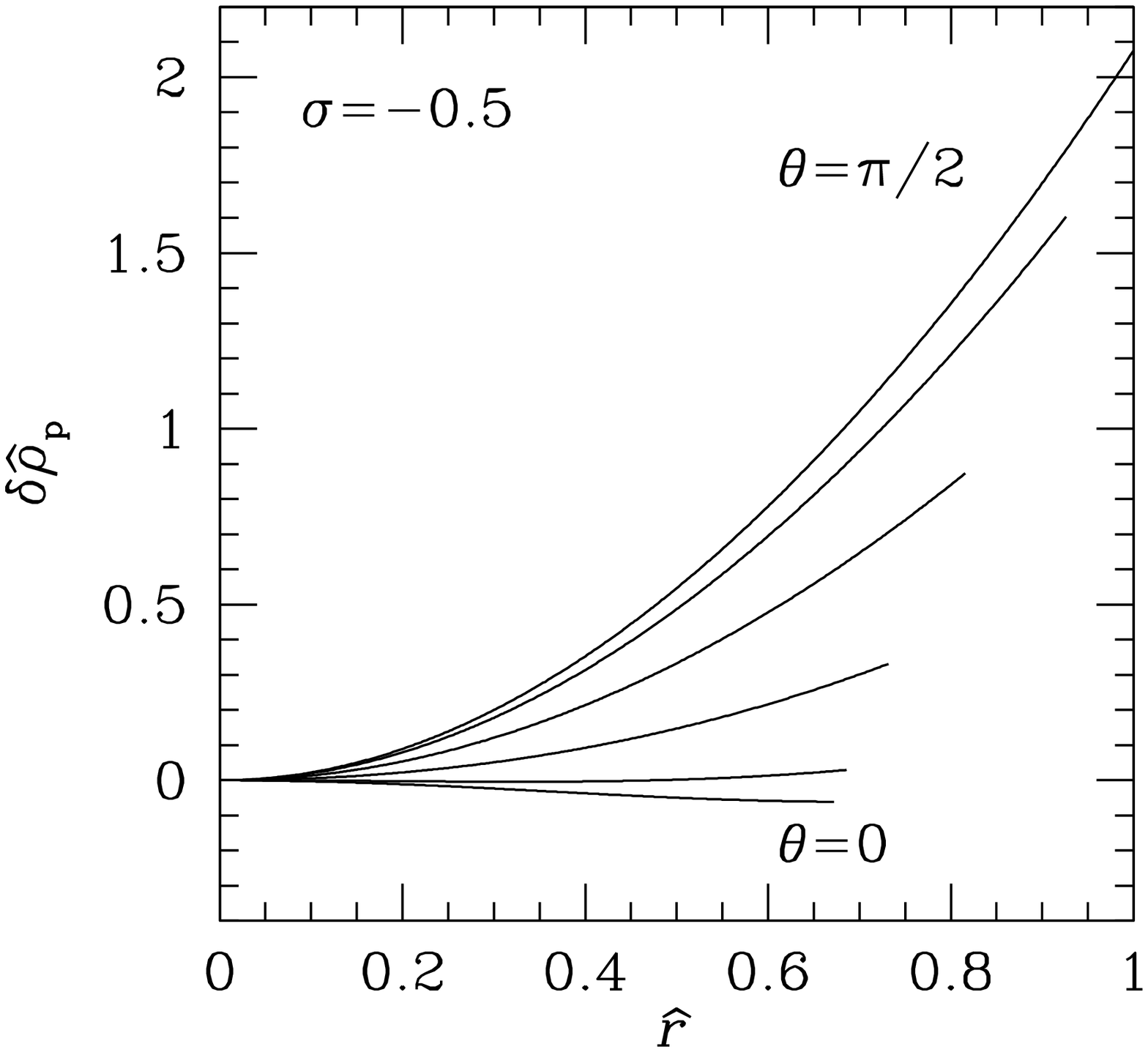}
\caption{Perturbed densities for the neutron (left) and the proton (right)
are given as functions of $\hat r$ for six different values of $\theta$. 
The model parameters are $\bar\Omega=0.8032$, $\sigma=-0.5$, and 
$(\delta\Omega_n,\delta\Omega_p)=(0,1)$.}
\end{figure}
\begin{figure}
\centering
\includegraphics[width=8cm]{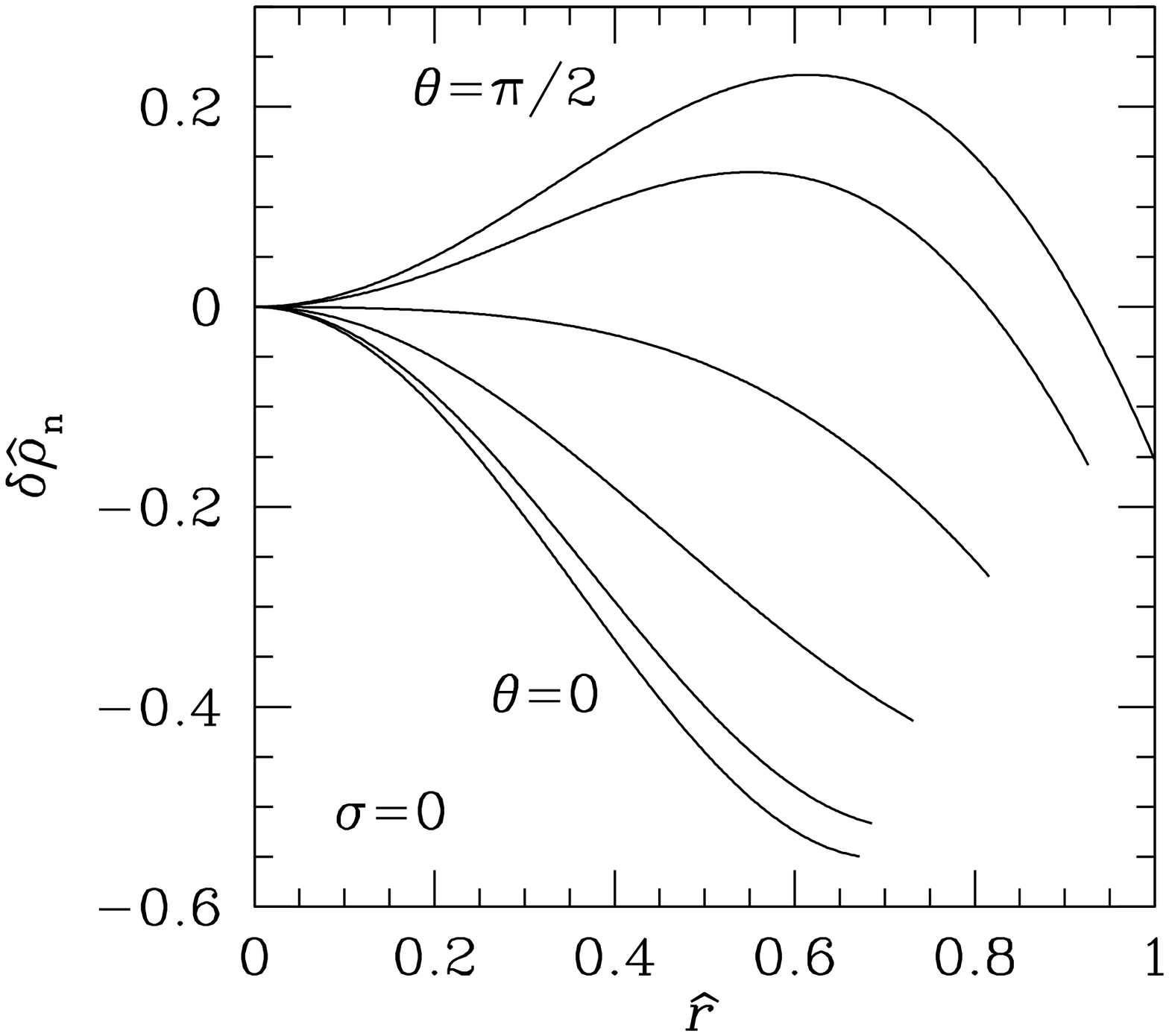}
\includegraphics[width=8cm]{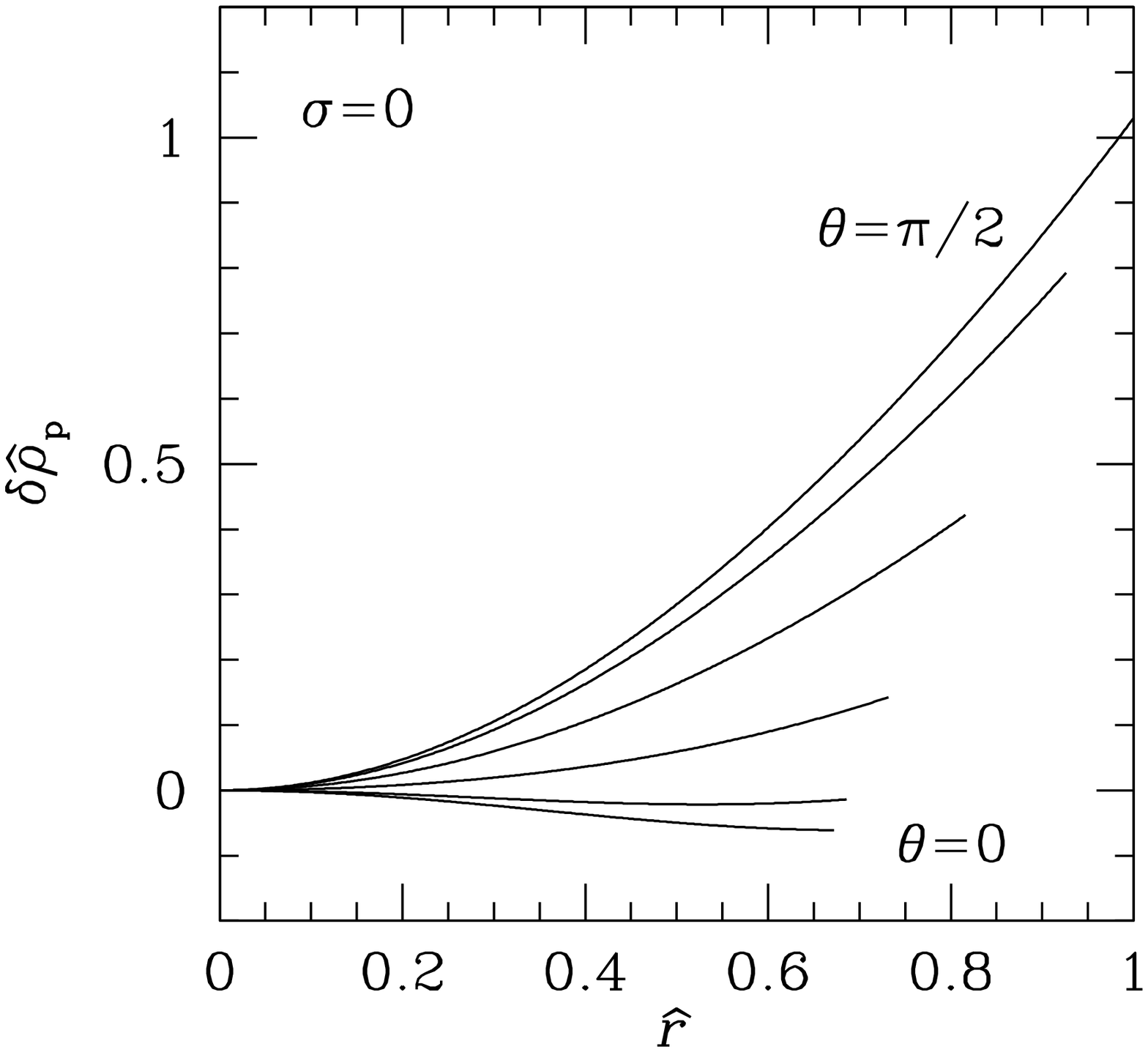}
\caption{Same as Figure 9 but for $\sigma=0$.}
\end{figure}
\begin{figure}
\centering
\includegraphics[width=8cm]{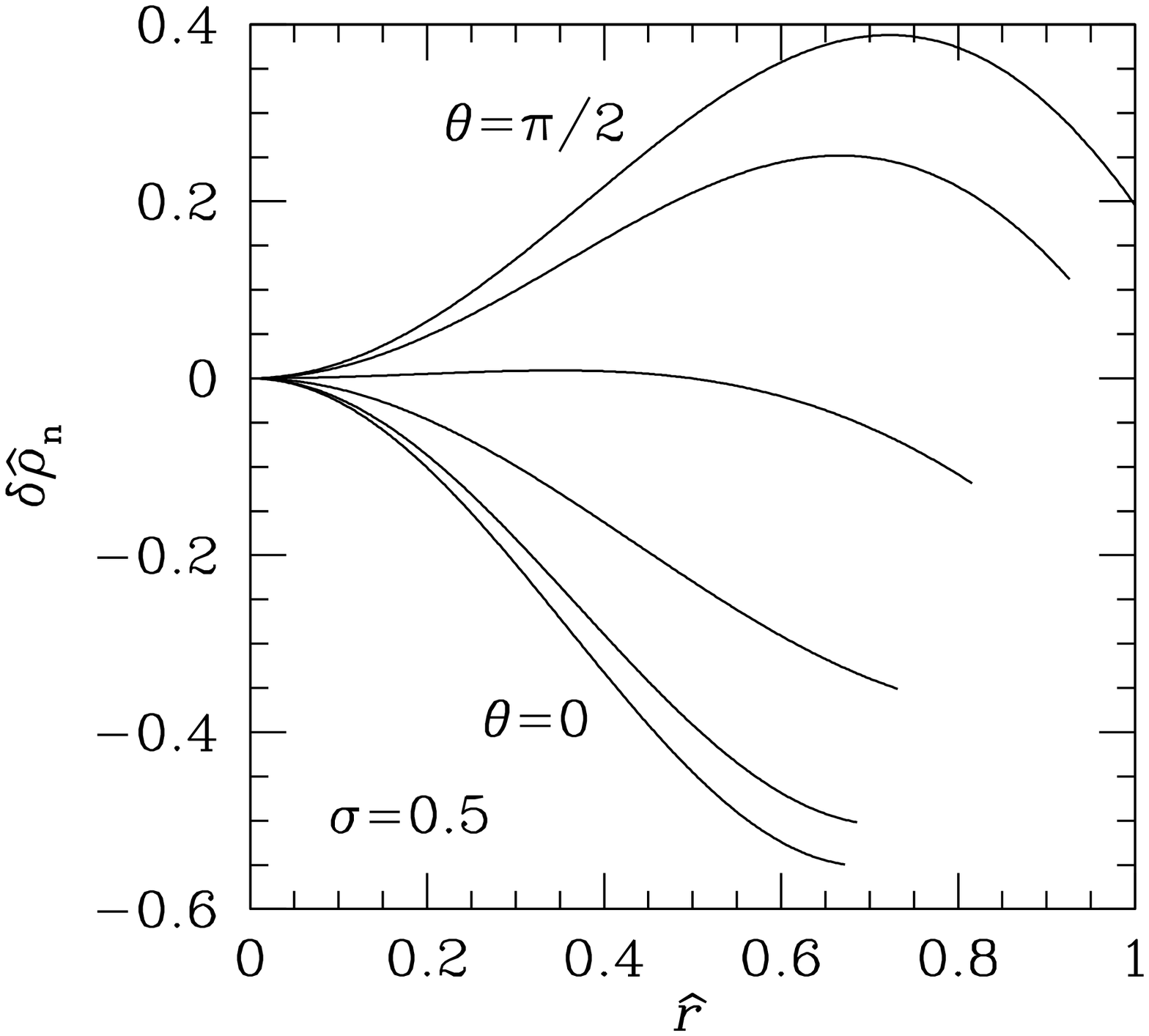}
\includegraphics[width=8cm]{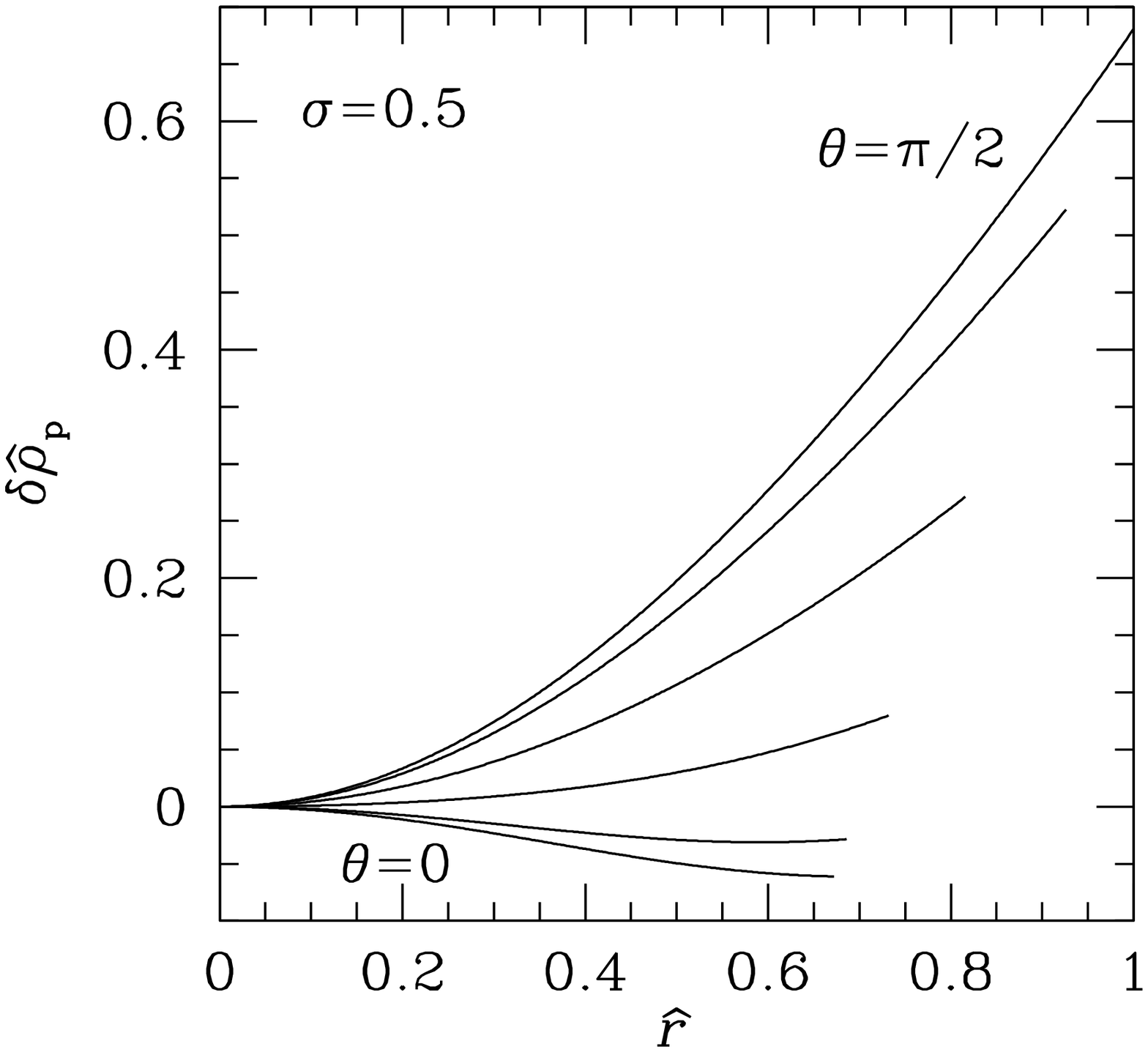}
\caption{Same as Figure 10 but for $\sigma=0.5$.}
\end{figure}
\begin{figure}
\centering
\includegraphics[width=8cm]{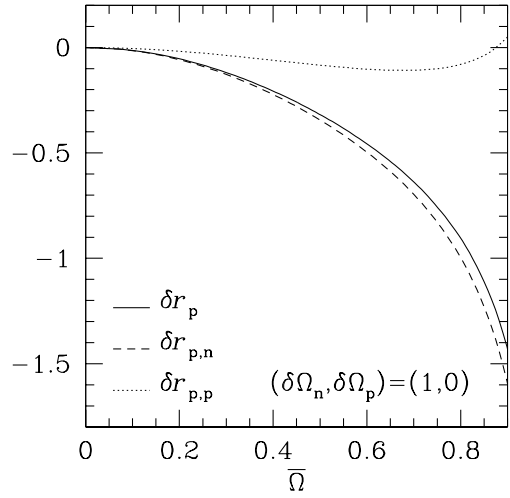}
\caption{Perturbed axis ratios $\delta r_p$, $\delta r_{p,n}$, and 
$\delta r_{p,p}$ are given as functions of $\bar\Omega$. The solid, dashed, 
and dotted curves denote $\delta r_p$, $\delta r_{p,n}$, and $\delta r_{p,p}$, 
respectively. The model parameter is $(\delta\Omega_n,\delta\Omega_p)=(1,0)$.}
\end{figure}
\begin{figure}
\centering
\includegraphics[width=8cm]{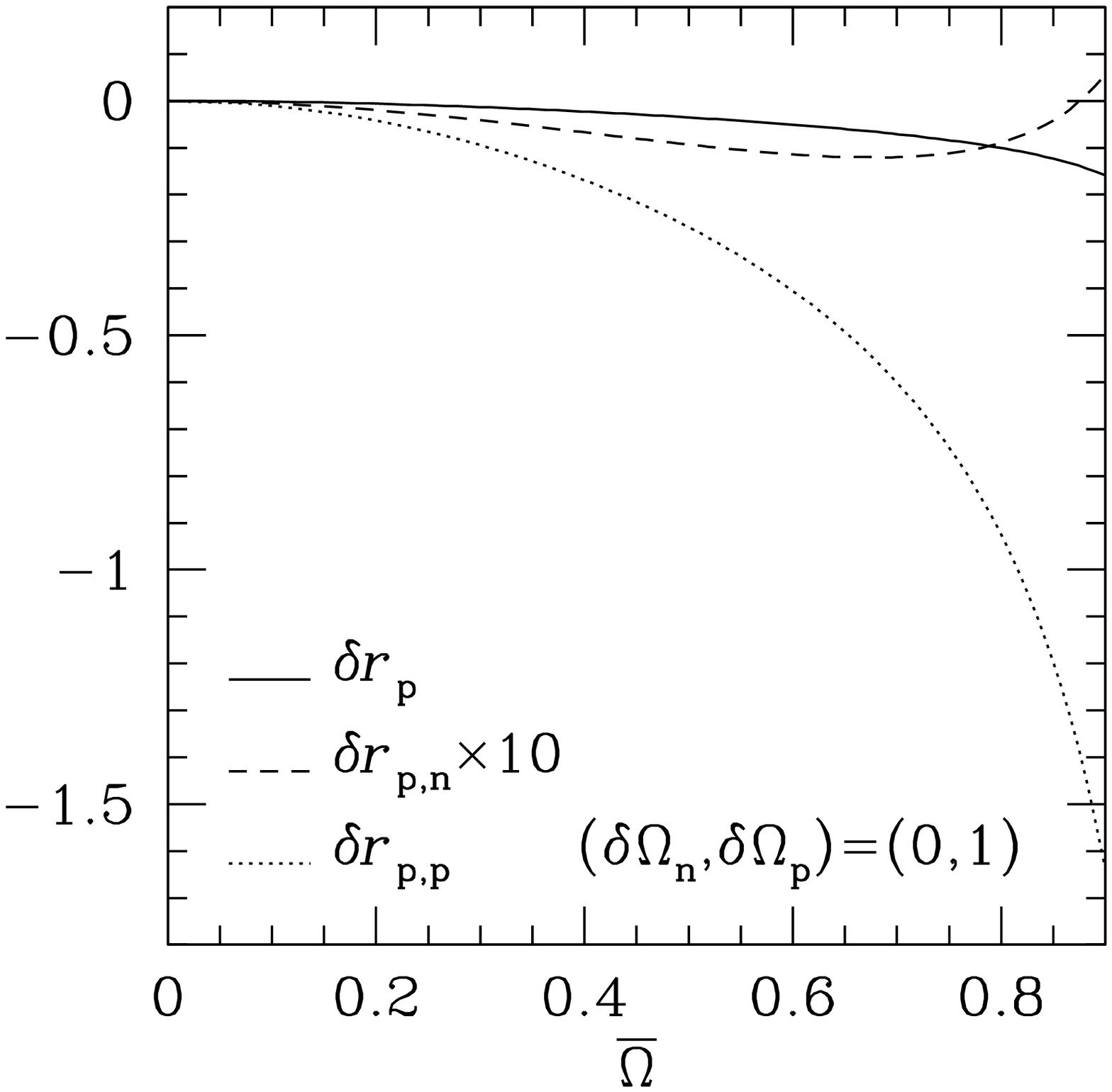}
\caption{Same as Figure 12 but for $(\delta\Omega_n,\delta\Omega_p)=(0,1)$.}
\end{figure}
\begin{figure}
\centering
\includegraphics[width=8cm]{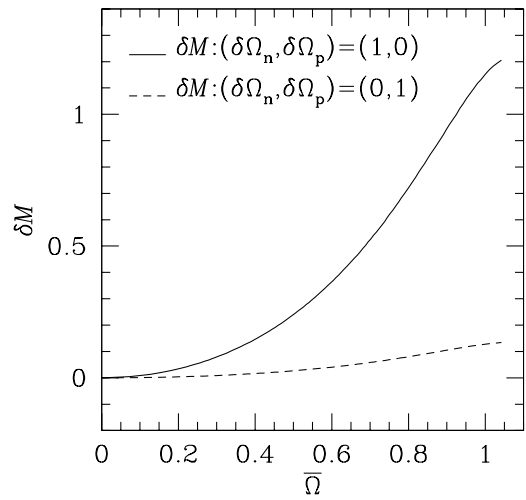}
\caption{Perturbed mass $\delta M$ is given as a function of $\bar\Omega$. 
The solid and dashed curves denote results for 
$(\delta\Omega_n,\delta\Omega_p)=(1,0)$ and 
$(\delta\Omega_n,\delta\Omega_p)=(0,1)$, respectively.}  
\end{figure}
\begin{figure}
\centering
\includegraphics[width=8cm]{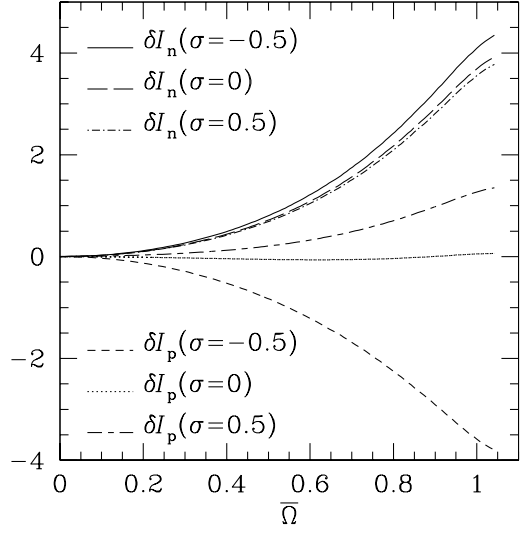}
\caption{Perturbed moments of inertia for the neutron and the proton 
$\delta I_n$, $\delta I_p$ are given as functions of $\bar\Omega$. 
The model parameter is $(\delta\Omega_n,\delta\Omega_p)=(1,0)$.}
\end{figure}
\begin{figure}
\centering
\includegraphics[width=8cm]{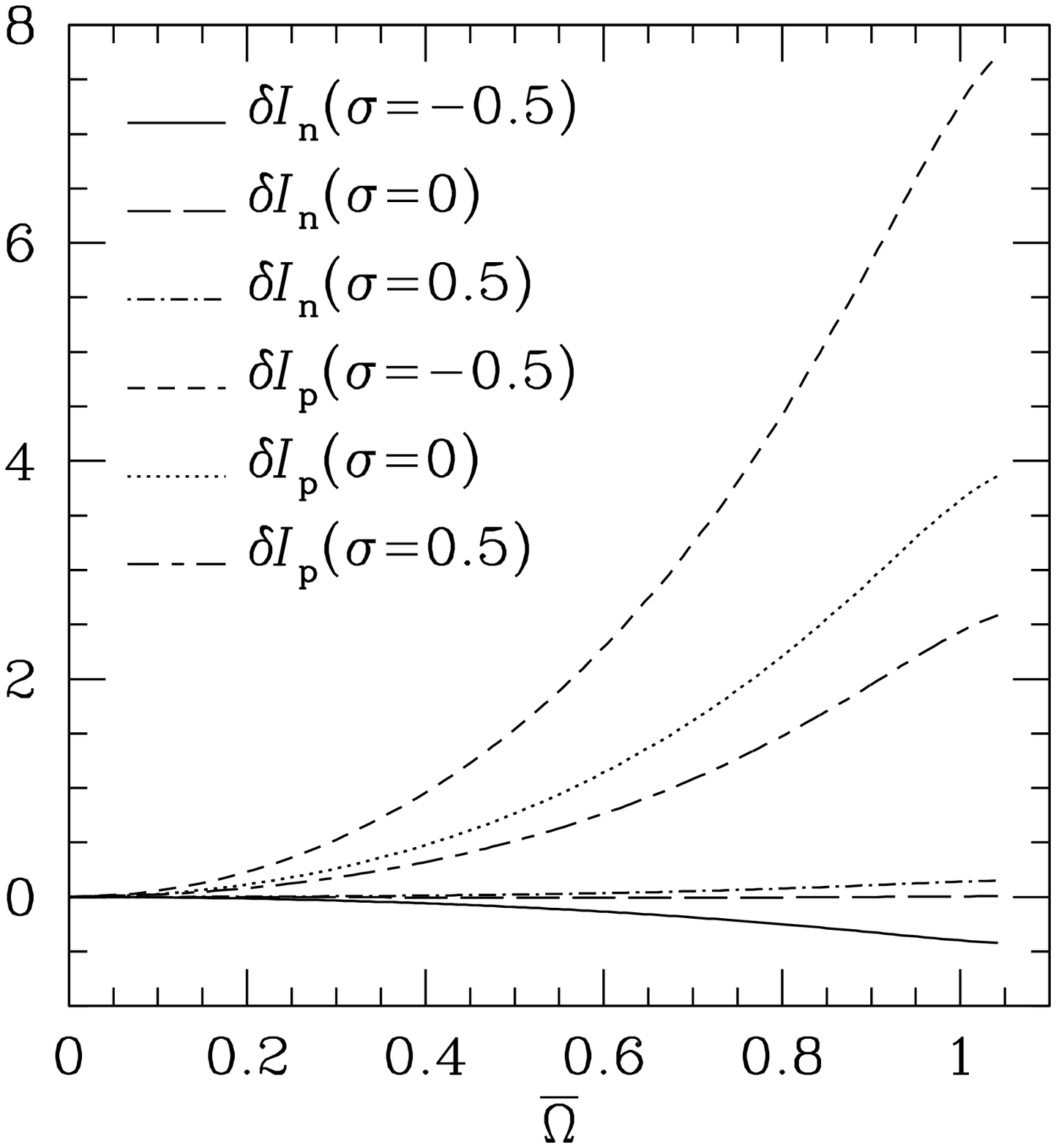}
\caption{Same as Figure 15 but for $(\delta\Omega_n,\delta\Omega_p)=(0,1)$.}
\end{figure}

\end{document}